# Imaging material functionality through 3D nanoscale tracking of energy flow
*Milan Delor[1], Hannah L. Weaver[2], QinQin Yu[2], Naomi S. Ginsberg[1,2,3,4,5]\**

Department of Chemistry[1] and Department of Physics[2], University of California Berkeley, Berkeley, California 94720, United States
Kavli Energy NanoSciences Institute[3], Berkeley, California 94720, United States
Material Sciences Division[4], and Molecular Biophysics and Integrated Bioimaging Division[5], Lawrence Berkeley National Laboratory, Berkeley, California 94720, United States
*nsginsberg@berkeley.edu

**The ability of energy carriers to move between atoms and molecules underlies biochemical and material function. Understanding and controlling energy flow, however, requires observing it on ultrasmall and ultrafast spatiotemporal scales, where energetic and structural roadblocks dictate the fate of energy carriers. Here we developed a non-invasive optical scheme that leverages non-resonant interferometric scattering to track tiny changes in material polarizability created by energy carriers. We thus map evolving energy carrier distributions in four dimensions of spacetime with few-nanometer lateral precision and directly correlate to material morphology. We visualize exciton, charge, and heat transport in polyacene, silicon and perovskite semiconductors and elucidate how disorder affects energy flow in 3D. For example, we show that morphological boundaries in polycrystalline metal halide perovskites possess lateral- and depth-dependent resistivities, blocking lateral transport for surface but not bulk carriers. We furthermore reveal strategies to interpret energy transport in disordered environments that will direct the design of defect-tolerant materials for the semiconductor industry of tomorrow.**

Energy flow is central to all biological, chemical and material functionality. Elucidating how distinct macroscopic functions emerge from different structural arrangements of atoms and molecules requires understanding how energy is transduced and transported between a system's building blocks. Materials science is undergoing a revolution with a burst of new high-performing semiconductors made from a diversity of readily tunable molecular building blocks[1–3]. Nevertheless, a fundamental understanding of why some semiconductors outperform others remains elusive[4], inhibiting rational materials design. The difficulty in gaining such predictive power is compounded by nanoscale spatio-energetic disorder, manifested in defects, impurities, and interfaces, that give rise to spatio-temporally heterogeneous energy transport. Elucidating how the microscopic details of a material relate to its emergent optoelectronic properties will therefore require the ability to individually and systematically correlate nanoscale structure to energy flow across a wide range of systems.

Resolving how energy flow is affected by spatio-energetic disorder requires tracking energy carriers over a wide spatiotemporal range – nanometers to microns and picoseconds to milliseconds – and directly correlating the measurement to material morphology. This correlation notwithstanding, powerful spatiotemporally-resolved approaches have recently been developed to visualize nanoscale energy flow using photoluminescence[5–7], transient absorption[8–11], or electron scattering[12] as contrast mechanisms. They achieve femtosecond time resolution[10,12], large dynamic range[5], chemical specificity[5–7,10], and excellent spatial sensitivity[5–7]. Nevertheless, their current implementations mostly track only a subset of energy carrier types, and rely on samples having specific optical or electronic properties, such as being absorbing yet low optical density[10], having large Stokes shifts[6] and appreciable emission[5,7], or being resistant to electron beams[12]. They also currently measure energy flow in two dimensions only, and most acquire a single image pixel at a time. These constraints limit the breadth of samples and fundamental processes that can be studied. Overcoming these challenges, we developed an approach that leverages elastic scattering, a universal optical interaction, to track evolving distributions of any type of energy carrier in 3 spatial dimensions, irrespective of their optical properties, as they move through any material on picosecond to millisecond timescales. Importantly, this approach enables simultaneous imaging of the nano-to-microscale morphological features that define the spatioenergetic landscape of the material, providing the much sought after *in situ* structure-function correlations.

In our approach, we first introduce a diffraction-limited, short pump light pulse to generate a localized collection of energetic carriers. These carriers act as point scatterers in the sample by modifying local material electric polarizability. We subsequently probe the pump-induced changes to the scattering profile of the material at controllable time delays over a large sample area through which energy carriers diffuse (**Figure 1a and Supplementary Sections 1-2**), thus imaging the evolving carrier distribution in space. Though related to transient absorption microscopy, the two key advances in our approach that help overcome limitations of previous methods are to operate in a reflection geometry and to image large areas using a widefield probe. Beyond providing advantages related to sample generalizability and contrast (**Supplementary Section 2.2**), a reflection geometry enables tracking energy carriers in 3D by interfering a reflected reference field (e.g. from the sample–substrate interface) with the backscattered field arising from the probe interaction with the energy carriers and their immediate surroundings. The resulting interferometric image converts phase delays (due to the varying depth of the scatterer distribution compared to the well-defined interface) into amplitude contrast, providing a semi-quantitative depth profile of the carrier distribution. A large probe area considerably reduces acquisition time by acquiring a snapshot of the sample's scattering profile in a single image exposure, obviating the need for sample, beam or detector scanning, and providing morphological correlation with diffraction-limited resolution. The signal can also be spectrally-resolved in the same instrument (**Supplementary Section 1**), a technique known as spectral interferometry[13–15]. The latter does not track carrier motion through space but instead provides information on the sample's spectral response to photoexcitation in localized regions of its heterogeneous landscape, a powerful tool when combined with the imaging mode of the instrument.

Inspired by advances in scattering-based interferometric and photothermal microscopies[16–23] that achieve exquisitely sensitive detection of tiny scatterers down to single molecules, we call our approach stroboscopic scattering microscopy (stroboSCAT). Whereas interferometric scattering microscopy (iSCAT) allows single-particle tracking of nanoparticles and biomolecules, stroboSCAT opens a new range of possibilities to track spatiotemporal evolution of distributions of photogenerated energy carriers. We use iSCAT's formalism to treat the pump-induced contrast changes (**Supplementary Section 2**); a similar formalism based on the optical theorem has been used for transient spectral interferometry of quantum dots[13,14]. Microscopically, both absorption and reflection originate from scattering; macroscopically, however, in the visible range most semiconductors possess refractive indices that are dominated by their relatively wavelength-independent real part ($n$) rather than imaginary part ($k$, responsible for absorption). For these high $n/k$ ratios, stroboSCAT is primarily sensitive to changes in $n$ [24–26] (**Supplementary Section 2.2**). Overall, stroboSCAT is a high-throughput method (<10 minute acquisition for full spatiotemporally resolved datasets) to track time-evolving carrier distributions in 3D with high spatiotemporal resolution and in-situ morphological correlation in a wide range of materials (opaque or transparent, emissive or not, composite or uniform). We show below that these distinctive attributes enable detailed and model-free structure-function correlations in a broad range of materials, precisely pinpointing the origin of functional heterogeneity in disordered semiconductors.

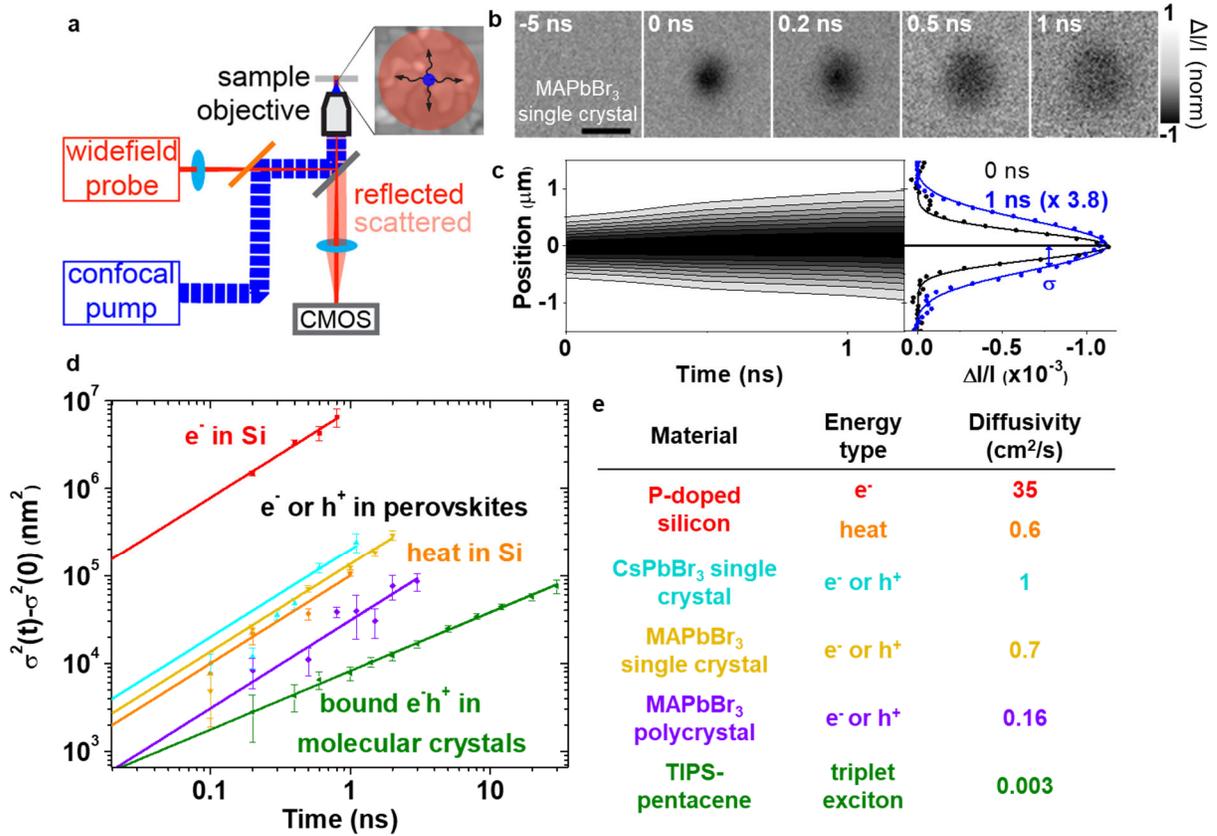

**Figure 1. Visualizing semiconductor exciton, charge and heat transport across four orders of magnitude in space and time**. (a) stroboSCAT setup. A confocal pump (306 nm diameter, λ = 440 nm) and widefield probe (8 μm diameter, λ = 635 nm) are overlapped in the sample. Both probe light scattered by the sample and reflected at the sample-substrate interface are imaged on the camera (CMOS). Full details in **Supplementary Sections 1-6.** (b) Example stroboSCAT dataset for a MAPbBr$_3$ single crystal, showing charge carriers diffusing as a function of pump-probe time delay. The peak pump-injected carrier density is 2 x 10$^{18}$ cm$^{-3}$ and the probe is spectrally far from the band edge (570 nm). The peak power densities at the sample are on the order of 0.2 MW/cm$^2$, far below the onset of nonlinear optical effects. All stroboSCAT plots are generated by taking the difference between pump$_{ON}$ and pump$_{OFF}$ raw images, normalized to the raw pump$_{OFF}$ image. Scale bar 1 μm. The spatiotemporal population distribution along the horizontal spatial axis is plotted in c, along with 1D Gaussian profiles extracted at 0 and 1 ns pump-probe delay. (d) Charge, exciton, or heat distributions vs time measured for a range of semiconductors (**Supplementary Section 7**). Error bars represent the 95% confidence intervals from Gaussian fits. All experiments are performed in a linear excitation regime, as confirmed by a lack of power dependence to extracted diffusivities. (e) Diffusivities extracted from linear fits of the data in (d).

We use stroboSCAT to visualize energy flow in a wide range of semiconductors, demonstrating its capability over four orders of magnitude in space and time, on both neutral and charged excitations migrating through organic, organic-inorganic, and inorganic semiconductors. Before focusing on disordered semiconductors, we illustrate stroboSCAT imaging of carrier diffusion in an ordered one:

**Figures 1b,c** display the spatial profile of charge carriers as a function of pump-probe delay in a methylammonium lead bromide (MAPbBr$_3$) perovskite single crystal. The probe at 635 nm is spectrally far from the band edge (570 nm), primarily detecting changes to *n*. Using ultra-stable picosecond pulsed laser diodes, we achieve shot-noise-limited differential contrast with sensitivities approaching 10$^{-5}$ and a signal-to-noise ratio averaging 40 for up to 1 ns pump-probe time delay with less than 1-minute integration per time delay. In this simple example, the diffusivity *D* for the charge carriers can be modelled from the Gaussian distribution variance of the scattering profile with time, $2Dt = \sigma^2(t) - \sigma^2(0)$ (**Supplementary Section 3**). The achievable sample-dependent spatial precision, $\Delta\sigma(t) = \pm 2\text{-}10$ nm for a <1 minute measurement per time delay (or $\Delta\sqrt{2Dt} = \pm 4\text{-}20$ nm), is not limited by diffraction but rather by fitting precision, which depends on signal-to-noise ratio[5,10]. **Figures 1d,e** summarize similar analyses on a variety of semiconductors using the same setup configuration to image heat, neutral bound pairs of charges (excitons), and free charge carrier diffusion. Because each type of energy carrier changes the local polarizability differently, the magnitude and sign of the stroboSCAT contrast permits distinguishing between types of energy carriers that may co-exist. For example, heat can generate opposite contrast to that of free charges, as we see in silicon (**Figure S4**).

Our results closely match published values for materials whose energy diffusivities have previously been determined[27–30], confirming stroboSCAT's viability. One notable observation in **Figures 1d,e** is that carrier diffusivities in MAPbBr$_3$ are reduced more than fourfold in disordered polycrystalline films compared to single crystals. Nevertheless, the impact of domain interfaces on energy flow can be non-trivial, depending greatly on their type, size, distribution and composition. Below, we use stroboSCAT to track and morphologically correlate energy flow up to, within and across energetic obstacles, using two classes of emerging semiconductors as case studies. Our results and corroborating simulations reveal that carrier trajectories are governed by highly anisotropic paths of least resistance, precluding the viability of diffusive models extracted from bulk or averaged measurements and calling for new ways to interpret and quantify energy flow in disordered environments.

We first explore the effect of low-curvature domain interfaces on exciton migration in polycrystalline 6,13-bis(triisopropylsilylethynyl)pentacene (TIPS-Pn) films. TIPS-Pn is a promising singlet-fission sensitizer for hybrid solar panels and represents an archetypal system to study energy transport in π-stacked molecular crystals (**Figure 2a**)[31,32]. In **Figure 2b**, iSCAT images at two different probe field polarizations display two orthogonally-oriented crystalline domains (light vs dark) separated by straight interfaces. Exciton migration imaged by stroboSCAT within a domain (**Figure 2c-e**) shows that at early time delays, the migration is non-linear, lending support to previous reports that attribute this behavior to the interchange between fast-diffusing singlet and slow-diffusing triplet pair excitations[33]. Beyond 5 ns, the linear diffusion that we observe at $D=0.003$ cm$^2$/s (**Figure S12**) is consistent with the common interpretation of free triplet migration[5,33].

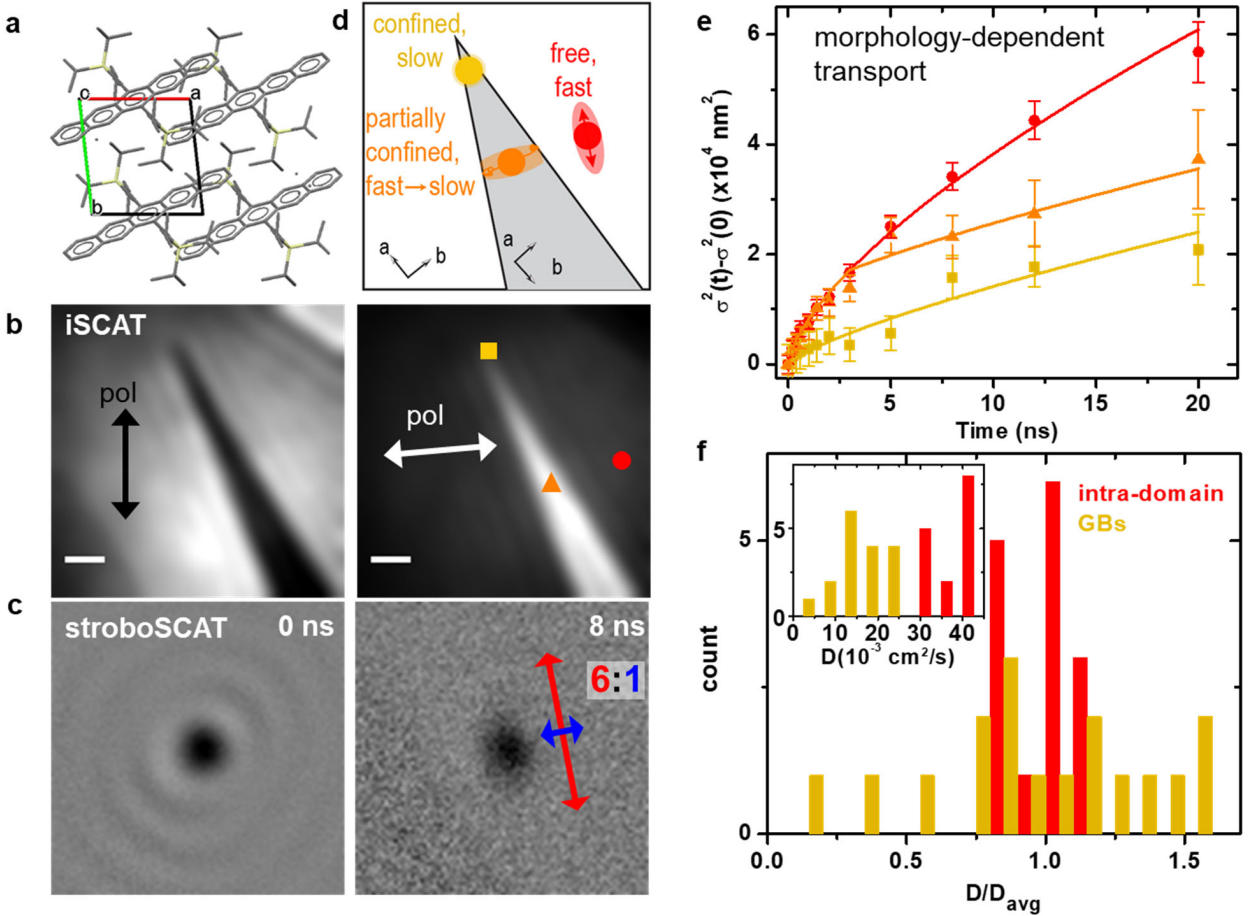

**Figure 2. Morphology-dependent exciton transport in TIPS-Pentacene.** (a) crystal structure of TIPS-Pn [34], displaying the orientation of the crystallographic *a* and *b* axes. (b) iSCAT images at different polarizations (pol) showing two orthogonally-oriented crystal domains. These images are bandpass-filtered to remove diffraction fringes near interfaces (**Figure S8**). (c) stroboSCAT images at 0 and 8 ns time delay in a crystalline domain (red spot in b), displaying anisotropic diffusion with six-fold faster transport along the π-stacked (red) axis of the crystal (**Figure S9**). The lab frame anisotropy directionality changes in different domains in relation to their crystal orientation. Both pump (440 nm) and probe (640 nm) are resonant with ground-state absorption in this example. (d) Schematic of exciton diffusion behavior at three different spots – intra-domain (red), confined (yellow), and intermediate transport scenarios (orange). (e) Corresponding population expansion dynamics along the fast diffusion axes. The pump is circularly polarized and the probe polarization is chosen to avoid contrast bias across domains (**Figure S10-S11**). Error bars represent the 95% confidence intervals from Gaussian fits. (f) Spot-to-spot variability of initial diffusivity $D_0$, determined from the fitting function $f(t)=2D_0t^\alpha$ with $\alpha$ being a free parameter, for intra-domain and GB-confined scenarios. Peak fluence is 140 μJ/cm$^2$ (**Figure S12**). Scale bars are all 1 μm.

The key finding enabled by stroboSCAT is the degree to which individual grain boundaries (GBs) hinder exciton transport in molecular crystals and that the extent of this hindrance varies widely. The narrow wedge-shaped central domain in this example provides an opportunity to systematically quantify the effect of domain confinement by interfaces on exciton migration (**Figure 2d**). Population distribution expansion along the fast migration axis at red, orange, and yellow spots highlighted in **Figures 2b,d** are plotted in

**Figure 2e**. Comparing the bulk crystalline domain (red) and the most confined spot (yellow), we find that interfaces severely hinder exciton transport, slowing it approximately four-fold. Transport at the partially-confined area (orange) can be accurately modeled piecewise, with free migration up to 3 ns (overlapping orange and red curves) and confined migration thereafter (orange curve parallels yellow curve), indicating a transition from bulk-like to slower transport when reaching an interface. We measured exciton migration in 15 different domains and at 17 different GBs and found that (i) exciton transport is always slower at GBs than in domains (**Figure 2f inset**); and (ii) transport speed is consistent in all measured domains, whereas transport at GBs is highly variable, as indicated by the normalized distribution of initial ($t=0$) diffusivities in **Figure 2f**. Indeed, interface formation kinetics, degree of lattice misorientations, and void and impurity concentrations will give rise to a wide range of transport behavior at different interfaces. Thus, high-throughput and correlative measurements of exciton migration over nanometer length scales provide the crucial ability to investigate energy transport properties *in situ* for each interface and surrounding crystal domains and to correlate these to their specific morphologies.

Although large crystalline domains separated by abrupt interfaces provide a systematic environment to test the effects of crystalline mismatch on energy transport, a more commonly-encountered morphology in polycrystalline semiconductors consists of sub to few-micron-sized domains. In these materials, energy carriers almost inevitably encounter domain boundaries during their lifetimes. GBs and, more generally, morphological boundaries (MBs) between domains (which are most typically crystal grains) thus significantly impact bulk-averaged measures of energy flow such as charge mobility and recombination[35,36]. There is, however, little consensus on the effect of MBs on the functional properties of a wide range of semiconductors. Nowhere is the debate currently more salient than with metal-halide perovskites[37–45]. For example, despite numerous studies suggesting MBs in perovskites have large trap densities, act as recombination centers, and impede carrier transport, photovoltaic efficiencies for polycrystalline films exceed 20% and do not necessarily scale favorably with grain size[46]. The primary difficulty in resolving these paradoxes lies in elucidating how the functional impacts of MBs locally deviate from bulk-averaged metrics. This challenge is exacerbated by the vast diversity of preparation protocols for polycrystalline metal halide perovskites, leading to radically different material properties[47]. We show below that visualizing carrier distributions in 3D as they trace the paths of least resistance through different halide perovskite films provides systematic and individualized detail on the effect of traps, the lateral- and depth-dependent conductive properties of MBs, and the resulting spatiotemporal anisotropy of charge transport as a function of material morphology.

The stroboSCAT time series in **Figure 3a** illustrates differences in charge carrier transport for three polycrystalline methylammonium lead iodide (MAPbI$_3$) films prepared using widespread preparation protocols that lead to different domain sizes (**Supplementary Section 4**). Because these samples are emissive, we also display the correlated steady-state widefield emission pattern arising from carrier recombination[40] in **Figure 3b** (**Supplementary Section 1**). The good correspondence between widefield emission and stroboSCAT images at late time delays confirms that the full extent of carrier migration is captured by stroboSCAT. On average, films with smaller domains exhibit slower lateral carrier transport, confirming that MBs negatively affect inter-domain carrier transport. From the data in **Figure 3a** we extract lateral diffusion lengths azimuthally- and time-averaged over the first 2 ns of 180 nm, 200 nm, and 700 nm for films made respectively with PbAC$_2$, PbI$_2$ and PbCl$_2$ precursors.

Importantly, in the largest-domain sample in **Figure 3a**, the sign of the stroboSCAT contrast can reverse from negative to positive. By correlating stroboSCAT measurements to structural maps in the same field of view (**Supplementary Section 8)**, we show that these contrast flips occur only at MBs. We rule out that the contrast flips arise from a change in carrier density, scattering amplitude or heat, as we do not observe these sign flips in any other region when varying the pump fluence over four orders of magnitude. We therefore attribute these contrast-flips to a change in the phase of the interferometric cross-term combining the reflected and scattered fields (**Supplementary Section 2**). The cross-term phase depends linearly on the depth of the scattering objects with respect to the sample-substrate interface, providing a measure of the distribution of scatterers along the optical axis[48,49]. The resulting stroboSCAT contrast is strongly negative for carriers located within ~30 nm of the interface and weakly positive for carriers located at depths of ~ 50 ± 20 nm (**Figure S21c**). Thus, the localized regions of positive contrast in these films indicate that, at MBs, the density of carriers at depths around 50 nm is significantly larger than within 30 nm of the surface. This observation suggests that when a carrier encounters a MB near the surface of the film, the path of least resistance leads deeper into the film rather than across the feature, resulting in a very low density of surface carriers at MBs. In contrast, subsurface carriers appear to cross MBs almost unimpeded. Our findings provide important mechanistic insight as to why conductive AFM on $MAPbI_3$ films indicated infinite MB resistance at the film surface but that, somehow, carriers still migrate to adjacent domains[43].

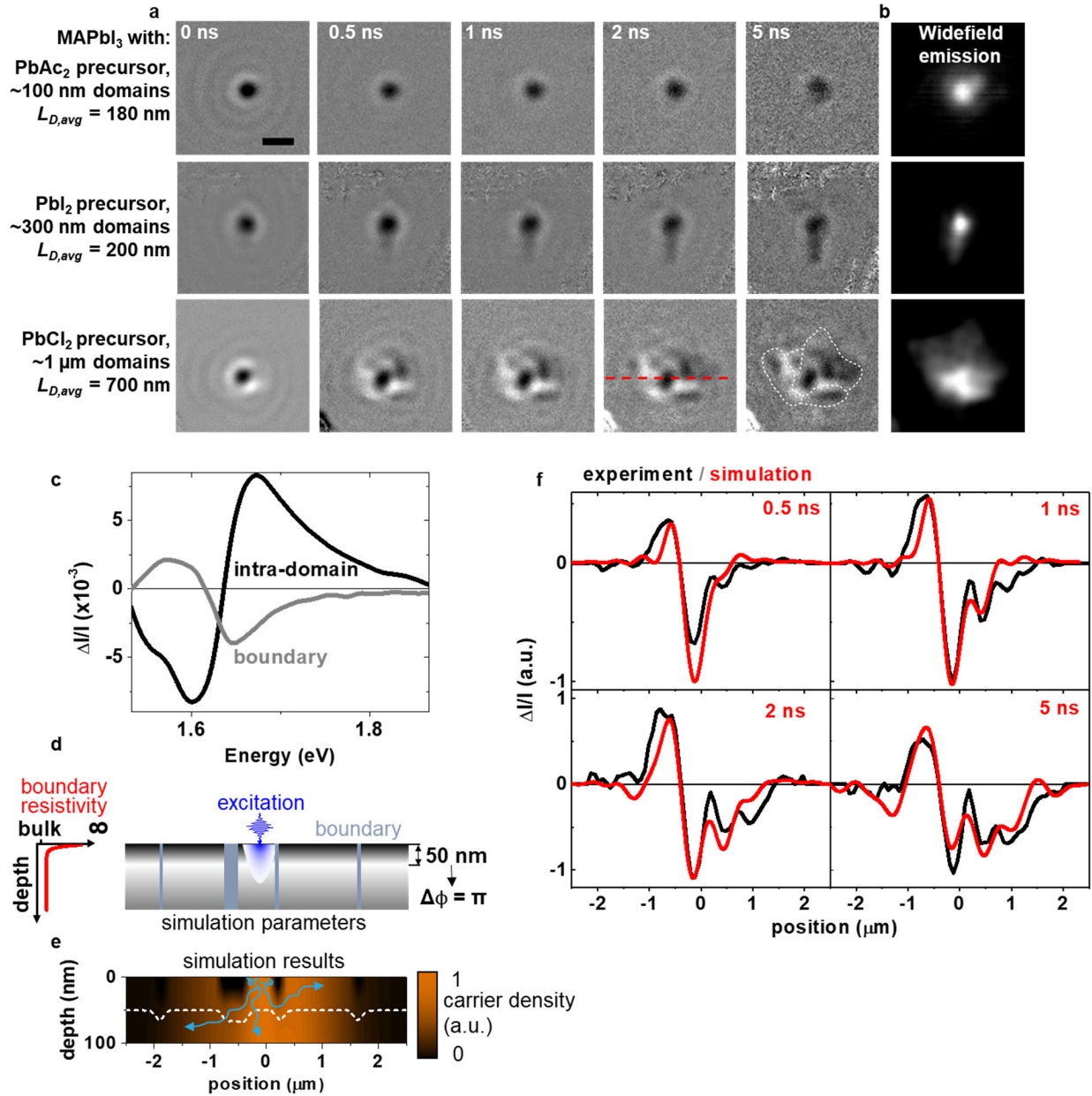

**Figure 3. Heterogeneous charge carrier transport in polycrystalline MAPbI$_3$.** (a) stroboSCAT time series on three MAPbI$_3$ samples prepared from different precursors, and (b) corresponding widefield emission images from confocal excitation. Domain sizes are measured using AFM (**Figure S13**). 2D diffusion lengths $L_{D,avg} = \sqrt{4D\tau}$ are azimuthally-averaged and are calculated over the first 2 ns only, during which $D$ is approximately constant. Normalized stroboSCAT contrast is represented in greyscale with black (white) representing the most negative (positive) value. The MB positions for the large-domain film are depicted using dashed-white lines. The signals do not depend on pump or probe polarization. Scale bar is 1 μm. Peak carrier density is 1.5 x 10$^{18}$ cm$^{-3}$. The probe at 640 nm is above the bandgap of MAPbI$_3$ but away from the band edge (760 nm) where the largest photoinduced absorption change occurs. (c) Spectrally-resolved interferometric signal near the MAPbI$_3$ band edge using the same excitation conditions but probing using broadband white light, showing a phase-flipped signal at MBs compared to within domains. (d) A sketch of a film cross-section used in simulations where MB positions were derived by comparing the simulations in (f) to the phase-sensitive stroboSCAT images for the PbCl$_2$ precursor film in (a). Greyscale shading represents the depth-dependent contrast expected from stroboSCAT measurements in MAPbI$_3$ (**Figure S21c**). (e) Results of simulations using time-propagated finite element analysis of carrier migration along the line cut indicated by the horizontal red line in (a), displaying the carrier distribution (orange shading) in the top 100 nm of the film 5 ns after excitation. The dashed white trace is the average axial position of the carrier distribution. Calculated distributions for other time delays are shown in **Figure S22**. The light blue traces are cartoons suggesting possible carrier trajectories that are consistent with the observed carrier distributions. (f) Quantitative agreement of experimental stroboSCAT data with simulation results after the simulation results have undergone appropriate contrast scaling and convolution with the apparatus' point spread function. The MB positions and resistivities used in the simulation to fit the experimental data are shown in panel (d). Additional details and datasets are provided in **Figures S14-S24**.

We corroborate our interpretation of the contrast flips using spectral interferometry (**Supplementary Section 1**): using near-diffraction-limited pump and probe beams, we measure the pump-induced spectral changes around the band edge of MAPbI$_3$. **Figure 3c** shows representative spectral response profiles within a domain (black trace) vs. at a boundary (grey trace) at $t$=0 for the large-domain MAPbI$_3$ film. The dispersive line shape within the domain corresponds closely to that reported in the literature from bulk transient reflectance spectroscopy of MAPbI$_3$ single crystals[25]. In contrast, the response from MBs occurs at the same spectral position but is inverted, indicating a ~π phase shift of the signal at the MB relative to within the domain. The conserved spectral position shows that the primary difference between intradomain and MB response is the phase, not the presence of another excited state species nor the chemical composition at the MB. Additional datasets (**Figures S19-S20**) show that intradomain signals always show the same dispersive lineshape, and that phase shifts between π/2 and π always occur at MBs. This trend, which persists across ~50 measured regions, confirms that charge carriers cross MBs only below the film surface in our samples.

We characterize depth-dependent MB resistivities by time-propagating finite element simulations of carrier diffusion in heterogeneous MAPbI$_3$ films with a simple model (details provided in **Supplementary Section 9**). We simulate the carrier distribution evolution in a 2D *x*-*z* film slice following an initial localized excitation (**Figure 3d**). To obtain reasonable agreement with experiment, the MBs must be parametrized by resistivities that are infinite at the film surface, concurring with results from AFM experiments[43], but that rapidly drop as a function of depth, approaching nearly intra-domain resistivity within ~50 nm of the film surface. **Figure 3e** displays the resulting carrier distribution after 5 ns. The dashed white trace represents the average axial position of the carrier distribution, indicating how the carrier density peaks further beneath the surface at the MBs. To directly compare the simulation with experiment and constrain our model, the simulated carrier distributions are contrast-scaled according to **Figure S21c** and convolved with the point spread function of the instrument. **Figure 3f** plots the resulting traces for four time delays, showing excellent agreement for these data and for additional structurally-correlated datasets shown in **Supplementary Section 9**. The combination of experiments and simulations thus enables a semi-quantitative description of 3D evolution of carrier distributions in these films: MBs act as impassable walls at the film surface, removing a lateral transport pathway for surface carriers. Carriers at depths below ~50 nm, however, cross almost unhindered into neighboring domains. Within domains, unimpeded 3D migration quickly leads to uniform carrier distributions. Since stroboSCAT is most sensitive to carriers located near the surface (**Figure S21c**), where the contrast is negative, we clearly distinguish domains (negative contrast) from MBs (positive contrast). This unique axial sensitivity afforded by phase contrast therefore provides a comprehensive 3D picture of both the morphological and functional properties of these materials. Furthermore, by measuring multiple regions and different films, we reveal that each MB has its own conductivity profile, leading to inter-domain energy carrier flow patterns that are highly anisotropic both axially and laterally (**Figures S15-16**).

To generalize our findings, we systematically quantify the degree of functional heterogeneity in these samples with further analyses. **Figure 4a** plots the initial angle-resolved charge carrier lateral diffusivity for MAPbI$_3$ prepared with the PbCl$_2$ precursor, which ranges from 0.1 cm$^2$/s to 1.1 cm$^2$/s. The time dependence of lateral carrier motion along each of the four color-coded directions in **Figure 4a** is depicted in **Figure 4b**. Although intra-domain transport both before and after passing through a MB can be as high as 1.3 cm$^2$/s, the terraces between the higher-slope portions of the curves observed in the time dependence in **Figure 4b** illustrate how MB encounters appear to temporarily halt lateral energy flow. Strikingly, the final plateau in population expansion occurs within several nanoseconds, indicating that a large fraction of carriers stop migrating on time scales much shorter than the average carrier recombination time (379 ns, **Figure S14**). This termination indicates that at these fluences the carrier density decreases on a few-ns timescale to below the material's trap-state density. Furthermore, it suggests that some traps do not act as recombination centers but may instead be hole and/or electron-selective. This analysis emphasizes the importance of tracking carriers until diffusion terminates, rather than extrapolating from the early-time constant diffusion through to the carrier lifetimes, which would falsely imply an average diffusion length in this sample of ~10 instead of ~1 μm. In total, we show that moving beyond averaged metrics in all spatiotemporal dimensions is essential to answer multiple prominent questions surrounding these materials and to address the functional impact of structural and electronic disorder. For example, we find that in our large-domain MAPbI$_3$ films, MBs (in many cases, GBs) do not act as recombination centers and only affect lateral transport significantly for surface carriers, not bulk carriers. For charge extraction in planar photovoltaic architectures, surface carriers are axially extracted and should therefore be minimally affected by MBs that are oriented perpendicular to the charge extraction layers.

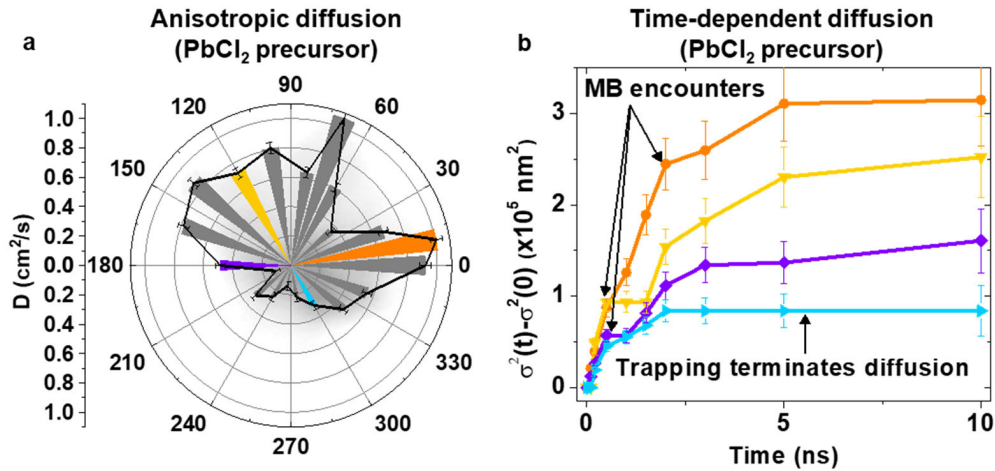

**Figure 4. Quantifying spatial and temporal carrier transport heterogeneity in polycrystalline MAPbI$_3$** (a) Angle-dependent diffusivities averaged over the first 2 ns for the PbCl$_2$ precursor data shown in **Figure 3a**. (b) Time-dependent diffusion for four representative azimuthal angles color-coded in panel (a). Error bars represent the standard deviation of the fits to several experimental datasets taken at the same location in the films. Additional analyses for films prepared from different precursors are presented in **Figure S18**.

Establishing the fundamental relationship between microscopic structural motifs and macroscopic function has been a longstanding multiscale challenge. In response, we devised a highly accessible and high-throughput strategy to measure energy flow *in situ* that is universally applicable to different energy forms and material properties. Benefitting from the high spatiotemporal resolution, sensitivity, and dynamic range of stroboSCAT, we demonstrated 3D measurement of energy flow through heterogeneous environments on pertinent scales, enabling direct correlation of the structure and functional connectivity in a broad range of semiconductors. We envision that stroboSCAT will impact the study of energy materials well beyond the scope of this present work, and will eventually reach the ultimate sensitivity limit of tracking single energy carriers[50], permitting nanometric functional mapping using single-particle localization with few-nm precision in all spatial dimensions. Finally, using scattering as a contrast mechanism enables a comprehensive range of processes - not only energy flow but also the transport of chemical species or ions - to be studied with the same level of detail and could shed light on catalytic cycles and chemical energy storage.

## Methods

Details of the experimental apparatus and sample preparation protocols are provided in the supplementary information.

Our implementation of stroboSCAT is illustrated in Figure S1. For all data shown in the text, the light sources used are two laser diodes (LDH-D-C-440 for the pump and LDH-D-C-640 for the probe, PicoQuant, with center wavelengths 440 and 635 nm, respectively) driven by the same laser driver. For the experiments shown in the text, we use a base laser repetition rate of 2 MHz, with the pump modulated at 660 Hz, and the pump-probe delay times are controlled using the electronic delay capabilities of the driver with 20 ps resolution. The temporal pulsewidths at FWHM are ~100 ps.

Both pump and probe are spatially-filtered through 20 µm pinholes and telescoped to ~6 mm and 1 mm beam diameters, respectively, before entering the microscope. The two beams are combined using a longpass filter (LP) and directed to a home-built microscope of very similar design to the microscope body detailed in Ortega Arroyo *et al.*[51] A f=300 mm wide-field lens (WFL) is inserted in the probe beam path upstream of the LP to focus the beam in the back focal plane of the objective, resulting in wide-field illumination (~5-60 μm depending on the beam size prior to the WFL) of the sample. A 50/50 beamsplitter (BS) reflects the pump and probe light into a high numerical aperture (1.4 NA) oil-immersion objective and onto the sample, resulting in an overlapped confocal and widefield illumination, respectively. Probe light reflected from the sample-substrate interface as well as scattered from the sample are collected through the same objective. The light transmitted through the beamsplitter is focused onto a charged metal-oxide semiconductor (CMOS) detector using a f=500 mm lens placed one tube length (200 mm) away from the back focal plane of the objective. The total magnification is 157.5. The pump light is spectrally filtered out in the stroboSCAT configuration. For widefield emission, other appropriate emission filter arrangements are used for any given sample. Optional half- or quarter-waveplates are used to control the polarization of pump and probe beams in polarization-sensitive measurements, such as for TIPS-pentacene.

For spectral interferometry measurements, the same event sequence, camera model and pump pulse excitation are used, but instead of using a narrowband probe, we use a broadband white light (WL) probe. The white light probe is generated by focusing the fundamental output (1030 nm, 200 kHz) of a Light Conversion PHAROS ultrafast regeneratively amplified laser system into a 3 mm yttrium aluminum garnet crystal. For the spectral interferometry data on MAPbI$_3$, the WL output is filtered with a 675 nm longpass filter to reduce sample exposure to above-bandgap light. The WL is sent collimated into the objective to obtain near-diffraction-limited probe pulses. The reflected light is then coupled into a home-built prism spectrometer and dispersed onto a CMOS camera. The entrance slit of the spectrometer is placed in the image plane. The electronic delays between pump and probe are controlled using an external delay generator, triggered with the pulse output of the ultrafast laser and feeding a user-delayed signal to the diode driver.

The use of electronic delays & modulation, as well as a widefield probe, results in no moving parts in the setup (apart from optional shutters), leading to an extremely stable and compact (<1 m$^2$) setup.

**References for Methods**
51.  Ortega Arroyo, J., Cole, D. & Kukura, P. Interferometric scattering microscopy and its combination with single-molecule fluorescence imaging. *Nat. Protoc.* **11**, 617–633 (2016).

**Acknowledgements:** This work has been supported by STROBE, A National Science Foundation Science & Technology Center under Grant No. DMR 1548924. Work toward manuscript revision was also supported by the 'Photonics at Thermodynamic Limits' Energy Frontier Research Center funded by the U.S. Department of Energy (DOE), Office of Science, Office of Basic Energy Sciences, under award DE-SC0019140. Q.Y. and H.L.W. each acknowledge a National Science Foundation Graduate Research Fellowship (DGE 1106400). N.S.G. acknowledges an Alfred P. Sloan Research Fellowship, a David and Lucile Packard Foundation Fellowship for Science and Engineering, and a Camille and Henry Dreyfus Teacher-Scholar Award.

**Author contributions:** MD designed and built the setup with QY. MD and HLW prepared samples and collected the data. MD analyzed the data. NSG supervised the research. MD and NSG wrote the manuscript with input from all authors.

**Competing interests:** The authors declare no competing interests.

**Data availability:** All raw data is displayed in Figures 1-3 of the main text and Figures S2- S24 of the supplementary data. Raw image files can be provided upon reasonable request to the authors.

Supplementary information for: **Imaging material functionality through 3D nanoscale tracking of energy flow**
*Milan Delor[1], Hannah L. Weaver[2], QinQin Yu[2], Naomi S. Ginsberg[1,2,3,4,5*]*
Department of Chemistry[1] and Department of Physics[2], University of California Berkeley, Berkeley, California 94720, United States
Kavli Energy NanoSciences Institute[3], Berkeley, California 94720, United States
Material Sciences Division[4], and Molecular Biophysics and Integrated Bioimaging Division[5], Lawrence Berkeley National Laboratory, Berkeley, California 94720, United States
*nsginsberg@berkeley.edu

1. **Detailed description of the stroboSCAT setup**

Our implementation of stroboSCAT is illustrated in Figure S1. For all data shown in the text, the light sources used are two laser diodes (LDH-D-C-440 for the pump and LDH-D-C-640 for the probe, PicoQuant, with center wavelengths 440 and 635 nm, respectively) driven by the same laser driver (PDL-828-S "SEPIA II" equipped with two SLM 828 driver modules and a SOM 828-D oscillator, PicoQuant). For the experiments shown in the text, we use a base laser repetition rate of 2 MHz, with the pump modulated at 660 Hz (every 3030 pulses), and the pump-probe delay times are controlled using the electronic delay capabilities of the driver with 20 ps resolution. We verified the calibration of the 'coarse' and 'fine' adjustments of the diode driver electronic delays using a computer-controlled mechanical translation stage (Newport) in a standard pump-probe geometry for delay times < 2 ns, and using an oscilloscope for delay times > 2 ns. We note that diodes and repetition rates are easily interchangeable for different experimental configurations. For short-time dynamics (<400 ps), diode afterpulsing can affect the accuracy of the measurements when using large diode powers. Therefore, care must be taken to minimize diode powers *at the source*, i.e. by reducing the driver current, rather than using neutral density filters in the beam paths. The resulting pulse FWHM are ~100 ps.

Both pump and probe are spatially-filtered through 20 µm pinholes (SF) and telescoped to ~6 mm and 1 mm beam diameters, respectively, before entering the microscope. The two beams are combined using a longpass filter (LP, DMLP505, Thorlabs) and directed to a home-built microscope of very similar design to the microscope body detailed in Ortega Arroyo *et al.*[1] A f=300 mm wide-field lens (WFL) is inserted in the probe beam path upstream of the LP to focus the beam in the back focal plane of the objective, resulting in wide-field illumination (~5-60 µm depending on the beam size prior to the WFL) of the sample. A 50/50 beamsplitter (BS) reflects the pump and probe light into a high numerical aperture (1.4 NA) oil-immersion objective (Leica HC PL APO 63x/1.40NA) and onto the sample, resulting in an overlapped confocal and widefield illumination, respectively. Probe light reflected from the sample-substrate interface as well as scattered from the sample are collected through the same objective. The light transmitted through the beamsplitter is focused onto a charged metal-oxide semiconductor (CMOS) detector (PixeLINK PL-D752, equipped with the Sony IMX 174 global shutter sensor) using a f=500 mm lens placed one tube length (200 mm) away from the back focal plane of the objective. The total magnification is 63 x 500 /200 = 157.5. On square pixels of 5.86 µm this magnification corresponds to 37.2 nm/pixel. One longpass (FEL550, Thorlabs) and one bandpass (FLH635-10, Thorlabs) filter are used to filter out pump light in the stroboSCAT configuration. For widefield emission, other appropriate emission filter arrangements are used for any given sample. Optional half- or quarter-waveplates are used to control the polarization of pump and probe beams in polarization-sensitive measurements, such as for TIPS-pentacene described in the main text. The aperture (Ap) below the beamsplitter in Figure S1 is used to switch between reflection-only mode and interferometric scattering mode to ensure the contrast is due to a change in the scattering cross-section (Figure S3). A 3 mm beam stop can also be inserted at the same position as Ap to switch to darkfield backscattering microscopy (Figure S3)[2–5].

For spectral interferometry measurements, the same event sequence, camera model and pump pulse excitation are used, but instead of using a narrowband probe, we use a broadband white light (WL) probe. The white light probe is generated by focusing the fundamental output (1030 nm, 200 kHz) of a Light Conversion PHAROS ultrafast regeneratively amplified laser system into a 3 mm yttrium aluminum garnet (YAG) crystal. For the spectral interferometry data on $MAPbI_3$, the WL output is filtered with a 675 nm longpass filter to reduce sample exposure to above-bandgap light. The WL is sent collimated into the objective to obtain near-diffraction-limited probe pulses. The reflected light is then coupled into a home-built prism spectrometer and dispersed onto a CMOS camera. The entrance slit of the spectrometer is placed in the image plane. This dual imaging/spectroscopy mode is similar to that recently implemented for widefield transient absorption microscopy[6]. The electronic delays between pump and probe are controlled using an external delay generator (DG 645, Stanford Research Systems), triggered with the pulse output of the ultrafast laser and feeding a user-delayed signal to the diode driver.

Although autofocusing capabilities as detailed in reference 1 were incorporated in the instrument, we found that stroboSCAT measurements are rapid enough and our microscope stable enough that autofocusing was not necessary, provided temperature fluctuations are minimal. Shutters are used to block pump and probe light (if desired) during program overheads to minimize sample exposure. The use of electronic delays & modulation, as well as a widefield probe, results in no moving parts in the setup (apart from optional shutters), leading to an extremely stable and compact (<1 $m^2$) setup, with system realignment needed only once every 2 months with daily use.

To trigger and synchronize the CMOS camera to the pump modulation rate, we use a phased dual-channel function generator to provide a 660 Hz TTL trigger signal to the CMOS and a synchronized 330 Hz trigger to the "Aux in" port of the laser driver. The latter starts each driver sequence that comprises 3030 pump pulses and 6060 probe pulses, both at a 2 MHz repetition rate. With the CMOS triggered at 660 Hz (and total exposure time set at 1.3 ms, averaging 2600 probe shots for each frame), consecutive images correspond to (1) probe with pump ON and (2) probe with pump OFF. The ratio pump ON/pump OFF for each consecutive pair is computed, and the ratio is averaged over 1000-3500 image pairs (total time with program overheads ~ 4-15 s per delay per scan for a full stroboSCAT image). Averaged pump OFF images (iSCAT) are simultaneously recorded at each time delay. After scanning through a full set of pump-probe delay times, the experiment's time delay scan is repeated $n$ times, with $n$ ranging from 3-10 depending on the signal to noise ratio of the measurement, with the final image for each delay corresponding to an average over $n$ time delay scans. To be able to record images at 660 Hz we reduce the region of interest to 192x192 pixels, i.e. ~ 7 x 7 μm, though faster cameras could allow even larger fields of view.

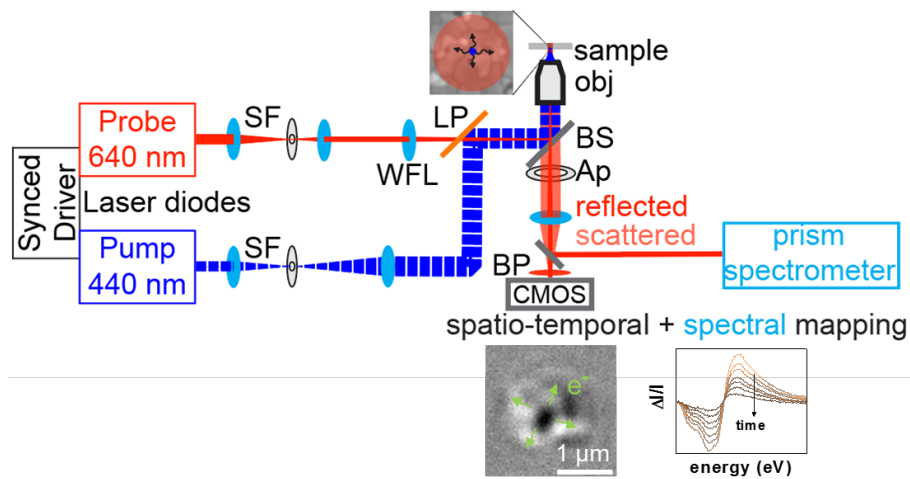

Figure S1. Schematic of the stroboSCAT setup. Laser diodes synchronized through a single oscillator/laser driver are used for pump and probe, using driver electronics to delay the probe with respect to the pump. After passing through spatial filters (SF), the beams are combined with a longpass filter (LP) before reflecting off a 50/50 beamsplitter (BS) into the objective (obj). The pump is sent collimated into the objective while the probe is focused in the back focal plane of the objective to result in confocal and widefield excitation spots, respectively. Light reflected from the sample-substrate interface or scattered from the sample is collected, spectrally filtered (through a longpass and bandpass filter, BP) and focused onto a detector (CMOS) for imaging. Alternatively, the collected light is sent to a home-built prism spectrometer with an entrance slit located in the image plane to perform spectral interferometry. An aperture (Ap) may be used block the scattered light and image primarily normally-reflected light. Alternatively, a 3 mm beam stop can be inserted instead of Ap to block normally-reflected light and only image scattered light[2,3].

The only image processing used is 2x2 pixel binning prior to Gaussian fitting (see Section 3 below). The stroboSCAT images shown in the text are raw, as-acquired images prior to 2x2 binning, and usually cropped to ~ 4 x 4 µm to emphasize the differential signal of interest. In principle, one could subtract negative-time delay stroboSCAT images for additional background subtraction, but we found this to have no effect on the fitting accuracy due to large signal-to-background ratio. Setup automation and data acquisition are implemented in LabVIEW 2014 64-bit. Data analysis and plotting is performed using a combination of imageJ (Fiji)[7,8], MATLAB and OriginPro.

The laser diodes provide high stability (<0.1% rms) and modularity in terms of repetition rates (single shot – 80 MHz), electronic delays (20 ps – 2 ms) and pulse sequencing, and fast warm-up times (<10 minutes from turning on) at the expense of time resolution compared to ultrafast lasers. We note that for interferometric contrast, the coherence length of the light source must be greater than the path difference between the scattered and reflected fields[1]. However, very large coherence lengths are undesirable as they can lead to interference among many optical elements, thus degrading image quality[9]. With a 2 nm spectral FWHM of the probe diode, we estimate a coherence length of ~40-60 µm depending on the medium's refractive index – much larger than the path difference for any films or crystals deposited on the substrate, but not large enough to lead to much interference from optical elements in the beam path. Ultrafast lasers with >30 nm spectral bandwidths (<2 µm coherence length) are in principle still coherent enough, thus providing the opportunity to extend stroboSCAT experiments to the femtosecond realm. The use of ultrashort pulses for pump excitation is not affected by coherence length constraints. Nevertheless, another advantage of using pulsed diodes is the low peak powers needed compared to highly impulsive (<100 fs) excitation for the same amount of overall excited state population. For short-pulse excitations, multi-photon effects, sample damage and heating must be carefully taken into consideration. For example, at the typical GW/cm$^2$ peak powers used in many ultrafast pump-probe experiments, carrier temperatures in semiconductors can reach ~$10^5$ K[10,11], leading to the observation of hot-carrier dynamics[11] over several hundred picoseconds that we did not observe in our experiments using peak powers that are three orders of magnitude lower.

As a general comment, we note that the concentration gradient created by a spatially-finite excitation does not influence the intrinsic energy carrier diffusivity: the only effect of a spatially-finite excitation is to allow us to observe diffusion because of a net flux of carriers down the concentration gradient. In our experiments, we keep excitation fluences below the onset of many-body effects, such that we can consider each carrier to act the same way that it would if there were only a single excitation in the material. Each carrier undergoes a random walk, which over time leads to a net flux out of the initial excitation spot.

## 2. stroboSCAT contrast mechanism
### 2.1. Interferometric scattering

In iSCAT, high sensitivity is achieved through interference of the scattered light of interest from the sample with light reflected at the substrate-sample interface (e.g. coverslip-sample interface). The light intensity reaching the detector, $I_d$, can be described as[1]:

$$I_d = E_i^2[r^2 + |s|^2 + 2r|s|\cos\phi]$$

where $E_i$ is the incident electric field, $r$ is the amplitude of the reflection coefficient at the interface, $s$ is amplitude of the scattering coefficient of the object of interest, and $\phi$ is the phase difference between scattered and reflected light. The scattering cross-section scales with the particle size raised to the third power, so the $|s|^2$ term, which is usually the signal of interest in dark-field microscopy, scales with particle size raised to the sixth power, contributing very little to the overall signal for scatterers <50 nm. However, the interferometric cross term scales with $|s|$ rather than $|s|^2$, and is amplified by the reflection term $r$. This term thus dominates for small particles and allows for extremely high sensitivity measurements to be made by simply increasing the incident electric field strength. Furthermore, $\phi$ can be expanded into[12,13]:

$$\phi = \phi_{gouy} + \phi_{scat} + \frac{4\pi z n}{\lambda}$$

where $\phi_{gouy}$ is the Gouy (focusing) phase difference between scattered and reflected light, which is constant (typically tuned to $-\pi \leq \phi_{gouy} < -\pi/2$ in iSCAT for maximum contrast[14], which we apply in stroboSCAT too) for a fixed objective-sample distance, $\phi_{scat}$ is the scattering phase (related to the material's complex refractive index; $\phi_{scat} \cong 0$ for off-resonant probing), $z$ is the object-interface distance, $n$ is the refractive index of the medium, and $\lambda$ is the illumination wavelength. Thus the relationship $\phi \propto z$ allows for three dimensional contrast. For the MAPbI$_3$ films in Figure 3 of the main text, $n(635$ nm$) = 2.6$, leading to a $\pi$ phase flip every 60 nm, which is similar to the pump and probe 1/$e$ penetration depths (50-70 nm).

In stroboSCAT, the differential contrast is defined as:

$$\frac{I_{pumpON} - I_{pumpOFF}}{I_{pumpOFF}} = \frac{I_{pumpON}}{I_{pumpOFF}} - 1 \approx \frac{2\cos\phi(|s_{pumpON}| - |s_{pumpOFF}|)}{r + 2|s_{pumpOFF}|\cos\phi}$$

assuming that $|s|^2 \ll r|s|\cos\phi$ and that $r$ does not change significantly between pump ON and pump OFF images (confirmed using the aperture Ap, Figures S1 and S3). Additional assumptions include that the ~-π Gouy phase difference between scattered and reflected light dominates the scattered phase contribution to $\phi$, such that the $\cos\phi$ term is virtually identical for the pump ON and pump OFF cases and is approximately equal to -1 for processes occurring at the sample surface. When carriers channel deeper into the sample, however, the depth-dependent pump ON contribution to the relative phase, $\cos(4\pi nz/\lambda)$, modulates the first term in the differential contrast expression above, enabling retrieval of depth-dependent information. The denominator can further be simplified to $r$, since $r \gg 2|s_{pumpOFF}|\cos\phi$. Thus, the contrast is proportional to the change in the scattering cross-section between the material in the presence vs. absence of excited state species, which in turn is directly proportional to the change in polarizability of the material between unpumped and pumped states. Overall, stroboSCAT benefits from the elegance, sensitivity and 3-dimensional contrast achievable with iSCAT, but expands it to the entirely different realm of ultrafast energy flow.

## 2.2. Distinguishing stroboSCAT from transient absorption microscopy

Although both absorption and reflection share a common origin in the scattered probe field, absorption is associated with the imaginary part of the refractive index ($k$), whereas reflection arises from changes to both $k$ and the real part of the refractive index $n$. For high-index materials, such as metal halide perovskites and the vast majority of inorganic semiconductors, $n$ is typically ~10-20 times larger than $k$ across the visible spectrum, so that to a good approximation, reflectivity is dominated by $n$ [15,16]. In the high $n/k$ limit, excited state reflectivity is also dominated by changes to $n$ [15–18]. The pure transient *absorption* spectrum (not transient transmission, which convolves both absorption and reflection) then corresponds to the Kramers-Kronig transform (i.e. a Hilbert transform) of the transient reflection spectrum[19]. Using spectral interferometry (i.e. transient reflection spectro-microscopy with a collinear, spatially well-defined reference field used for phase sensitivity), we observe a characteristic dispersive spectral profile around the band edge. This profile reproduces the transient absorption spectrum through a Hilbert transform. These data confirm that the signals we observe in reflection-based stroboSCAT for lead halide perovskites originate from changes to $n$, not $k$ **(Supplementary Section 8)**. In general, all of the data shown in the manuscript (except for TIPS-Pentacene) are taken with a probe that is spectrally away from band edge resonances, i.e. away from where the largest photoinduced absorption changes occur in semiconductors. Macroscopically, stroboCAT in these cases reports on changes to the real part of the refractive index. Microscopically, the signal arises from non-resonant Rayleigh scattering of the probe field with deeply subwavelength particles and their modified surroundings.

There are several advantages to probing in a reflection geometry:
  a. It is more generalizable than measurements in transmissive geometries since it does not require thin or optically transparent samples.
  b. The reference field (sometimes called local oscillator) refers to the unscattered part of the field which interferes with the scattered field to give rise to absorption/reflection. The reference field is much weaker in reflection than in transmission, since only a small fraction (typically 1-10%) of the forward propagating field is reflected. This results in a higher signal to background ratio, i.e. higher contrast, at the expense of lower photon counts, i.e. higher shot noise. Nevertheless, photoinduced changes to $n$ ($\Delta n$) display a relatively weak dependence on spectral detuning from the main absorption change at the band edge ($\Delta n \sim 1/\text{detuning}$, $\Delta k \sim 1/\text{detuning}^2$). The probe field can therefore be tuned away from absorptive wavelengths and into the transparency region of the semiconductor. Large probe powers at non-absorptive wavelengths are relatively benign and allow filling the well depth of the detector used, thus simultaneously maximizing contrast and minimizing shot noise.

c. In transmission-based geometries such as transient absorption, the co-propagating reference and signal fields that traverse the entire sample preclude extracting phase (depth) information over axial distances much smaller than the Rayleigh range without more sophisticated interferometric techniques such as off-axis holography. In contrast, a reflection-based geometry uses a spatially well-defined interfacial reflection as the reference field, which can provide up to few-nm sensitivity to depth.

d. When measuring in dense samples with finite dispersion and absorption (such as the >10 nm thick semiconductors measured here), forward-scattered light may undergo multiple scattering or re-absorption events prior to interfering with the local oscillator, quickly scrambling phase information. In stroboSCAT, by focusing on back-scattered light near the refractive index interface on the illuminated side, and remaining spectrally far away from large photoinduced absorption changes, we minimize phase scrambling.

e. Deeply subwavelength-sized objects have equal backscattering and forward-scattering contributions, whereas larger objects have much larger forward-scattering contributions.[20,21] In collinear third-order experiments where the signal field is necessarily emitted in the same direction as the reference field, detecting in epi geometry significantly enhances the ratio of signals arising from small objects near the refractive index interface (the signal of interest in stroboSCAT) vs. bulk contributions that can arise from birefringence, bulk heating, sample inhomogeneities etc.

Taken together, these fundamental differences give stroboSCAT a set of unique capabilities that are highly complementary to those of transient absorption: the ability to apply the same instrument to a vast range of different materials and track energy flow in a rapid, benign fashion in dense or opaque environments without changing wavelengths or detection geometry; the unique opportunity to extract 3D information through preserved phase sensitivity; and the potential to extract the full dielectric response of a material to photoexcitation without propagating assumptions through Kramers-Kronig transformations.

### 3. Data analysis

Following 2x2 binning of the raw stroboSCAT images, several strategies can be used to extract the diffusivities. For isotropic diffusion (e.g. perovskite single crystals), we plot the line profile for each time delay along any given axis, integrating across 4 bin-wide rectangular regions. The resulting profile is fitted with a Gaussian function for each time delay, as described in more detail further below. For anisotropic diffusion (e.g. TIPS-pentacene), the same strategy is used for both long and short diffusion axes.

In terms of achievable sensitivity, a typical <10-minute experiment consists of averaging 10,500 pumpON/pumpOFF image pairs per delay time. Given a camera well depth of ~ 30,000 electrons, following 2x2 pixel binning and taking 4-bin-wide line cuts, the total photoelectron count per image per 4-bin-wide area is ~ $5 \times 10^9$. In a shot-noise limited measurement, this provides a sensitivity floor of $1.4 \times 10^{-5}$. In the pump ON/pump OFF image, taking the ratio corresponds to a floor of $2 \times 10^{-5}$. For low-contrast samples, such as Si, longer averaging was used (10 scans), improving sensitivity to ~$10^{-5}$. Averaging over more pixels or integrating radially can also be done for homogeneous, fast-expanding signals like electrons in Si, further improving sensitivity to below the $10^{-5}$ level, if necessary. Near-shot-noise limited measurements were easily attained for homogeneous samples, while strongly scattering samples might include slightly higher noise levels due to imperfect background subtraction.

The Gaussian function used to fit normal spatial distributions is:
$$y(t) = A(t) * \exp\left(-\frac{(x-x_c)^2}{2\sigma^2(t)}\right)$$
where A(t) is a pre-exponential factor dependent on the contrast magnitude at each time delay $t$, $x_c$ is the center position, and $\sigma(t)$ is the Gaussian standard deviation for each time delay.

As detailed in Akselrod et al.[22], using the property that the variance of convolved Gaussians are additive, the solution to the diffusion equation in one dimension can be expressed as:
$$\langle x(t)^2 \rangle = \sigma^2(t) = \sigma^2(0) + 2Dt$$
where $\langle x(t)^2 \rangle$ is the mean square displacement, $\sigma^2$ is the variance of the population distribution at any given time, and $D$ is the diffusion coefficient. Thus, for ordered systems,
$$D = \frac{\sigma^2(t) - \sigma^2(0)}{2t}.$$

A more generalized form of the 1-dimensional diffusion equation applicable to both ordered and disordered systems can be written as:
$$\sigma^2(t) - \sigma^2(0) = 2D_0 t^\alpha$$
to account for subdiffusive transport behavior in disordered systems that exhibit a distribution of site energies and trapping, where $\alpha < 1$, or superdiffusive (ballistic) transport in systems where the time between species scattering events is long compared to observation times, where $\alpha > 1$.

We note that the variance obtained from Gaussian fitting of the intensity profile in stroboSCAT images is strictly speaking a convolution of the spatial distribution of excited species with the point spread function (PSF) of the system, which itself is a convolution of the individual pump and probe PSFs as well as that of the detector. However, since the latter three are invariant over time, they do not contribute to the difference signal $\sigma^2(t) - \sigma^2(0)$.

For heterogeneous, anisotropic diffusion, one possibility is to plot a radial profile averaged over all angles and fit it with a half-Gaussian to obtain an average intensity distribution. To obtain plots like in Figure 4a of the main text, the radial distribution is instead split into 10 to 30 degree slices and the intensity distribution of each slice is measured (the intensity at any given distance from the central point is the integrated intensity of the pixels over the arc bounded by the slice). However, when trapping occurs, for example at grain boundaries, the carrier population is not normally distributed and thus cannot be accurately fit with a Gaussian function. In these circumstances, we measure the distance from the center of the radial distribution at which the population drops to 1/$e$, and then convert the extracted half-width at 1/$e$ value ($w$) to a Gaussian standard deviation using the relationship $\sigma = \frac{w}{\sqrt{-2\ln(I_T)}}$, where $I_T$ is the chosen Gaussian intensity threshold (i.e. 1/$e$ in this case), assuming a normal distribution. Section 8 details our simulations of depth-dependent heterogeneous diffusion using finite element analysis, providing a more quantitative analysis of phase-sensitive stroboSCAT images in complex media.

4. **Sample preparation**

Sample substrates: all substrates are 22 x 22 mm or 24 x 50 mm VWR #1.5 glass coverslips. Every substrate is subjected to an extensive cleaning procedure as follows: 15 min sonication in a 2% Hellmanex solution in NANOpure deionized water, followed by several quick rinses in NANOpure deionized water; 15 min sonication in NANOpure deionized water; 10 min sonication in acetone; 10 min sonication in isopropyl alcohol; immediately dried under a flow of filtered nitrogen; and finally cleaned with an $O_2$ plasma for 3 minutes in a reactive ion etch chamber.

Reagents: All reagents were used as received without further purification. Methylammonium bromide (MABr, GreatCell Solar); methylammonium iodide (MAI, GreatCell Solar); Cesium bromide (CsBr, Alfa Aesar); lead bromide ($PbBr_2$, Alfa Aesar); lead iodide ($PbI_2$, Alfa Aesar); lead chloride ($PbCl_2$, Alfa Aesar); lead acetate trihydrate ($Pb(Ac)_2$, Sigma-Aldrich); 6,13-Bis(triisopropylsilylethynyl)pentacene (TIPS-Pentacene, Sigma-Aldrich); Trichloro(phenethyl)silane (TPS, Sigma-Aldrich); poly(methyl methacrylate) (PMMA, $M_w$=120,000, Sigma-Aldrich).
All solvents are purchased from Sigma-Aldrich.

Sample handling: All samples apart from Silicon and perovskite single crystals are prepared in a sealed glovebox with nitrogen atmosphere and with <2ppm $O_2$ and $H_2O$. Once prepared, the samples are sealed between two substrates using epoxy (EPO-TEK) in the glovebox to protect them from water and oxygen exposure during measurements. For atomic force microscopy measurements on thin films, another sample is prepared consecutively under the same conditions and using the same solution, but is not sealed between the two substrates and is measured immediately after preparation. Single crystals are all grown at ambient conditions. Once the crystals are grown, they are brought into the glovebox, placed on clean substrates, and 200 μL of a 40mg/mL solution of PMMA in chloroform is dropcast on the crystals to keep them in place and prevent exposure to oxygen and moisture during measurements.

Metal-halide perovskite single crystals: *$MAPbBr_3$ single crystals* were prepared according to a published procedure using antisolvent vapor diffusion[23]. Briefly, a 1:1 molar ratio, 0.2 M solution of $PbBr_2$/MABr was prepared in N,N-dimethyl formamide (DMF). The solution was then filtered using a 0.2 μm PTFE syringe filter. 2 mL were placed in a clean 4 mL vial, which was placed inside a larger scintillation vial filled with dichloromethane. The large vial was sealed and crystals were left to grow for 1 week, resulting in hyperrectangular crystals of dimensions ~ 3 x 3 x 1 mm.
*$CsPbBr_3$ single crystals* were grown using antisolvent vapor diffusion according to a published procedure[24]. The same procedure as that described for $MAPbBr_3$ was used, with a 1:1 molar ratio, 0.04 M solution of $PbBr_2$/CsBr in DMF and using nitromethane as antisolvent. The crystals were left to grow for 3 weeks, resulting in large hyperrectangular crystals of dimensions ~ 20 x 2 x 2 mm, which were cleaved before mounting on substrates for measurements.

$MAPbBr_3$ polycrystalline films were prepared by dissolving MABr and $Pb(Ac)_2$ in a 3:1 molar ratio in DMF with a final concentration of 0.5 M. The solution was spin-cast at 2000 rpm for 60 seconds. The films were subsequently annealed for 5 minutes at 100°C[25].

MAPbI$_3$ polycrystalline films were prepared using different published processing routes, described briefly below:

*Pb(Ac)$_2$ precursor films*[25] were prepared by dissolving MAI and Pb(Ac)$_2$ in a 3:1 molar ratio in DMF for a final concentration of 0.5 M. The solution was spin-cast at 2000 rpm for 60 seconds. The films were subsequently annealed for 5 minutes at 100°C.

*PbI$_2$ precursor films*[26] were prepared by dissolving MAI and PbI$_2$ at a 1:1 molar ratio at 200 mg/mL in DMF. The solution was spin-cast at 2000 rpm for 30 seconds. The films were subsequently annealed for 20 minutes at 100°C.

*PbCl$_2$ precursor films*[27] were prepared by dissolving MAI and PbCl$_2$ at a 3:1 molar ratio with final concentrations of 2.64 M and 0.88 M, respectively. The solution was spin-cast at 2000 rpm for 60 seconds. The films were subsequently left to dry for 30 minutes at room temperature in the glovebox, followed by annealing at 90°C for 150 minutes.

TIPS-pentacene was dissolved in toluene at 5mg/mL and filtered through a 0.45 µm PTFE filter. The substrates were treated by leaving to soak overnight in a petri dish with a solution of 190 µL TPS in 100 mL toluene, then rinsed with toluene and dried prior to deposition. The TIPS-pentacene solution was then spin-cast at 250 rpm, and then solvent-vapor annealed at 60°C in a toluene-saturated atmosphere for 24 hours.

Silicon wafers were prime-grade P-type, boron-doped wafers purchased from WaferNet Inc., used without further modification.

## 5. Current system resolution

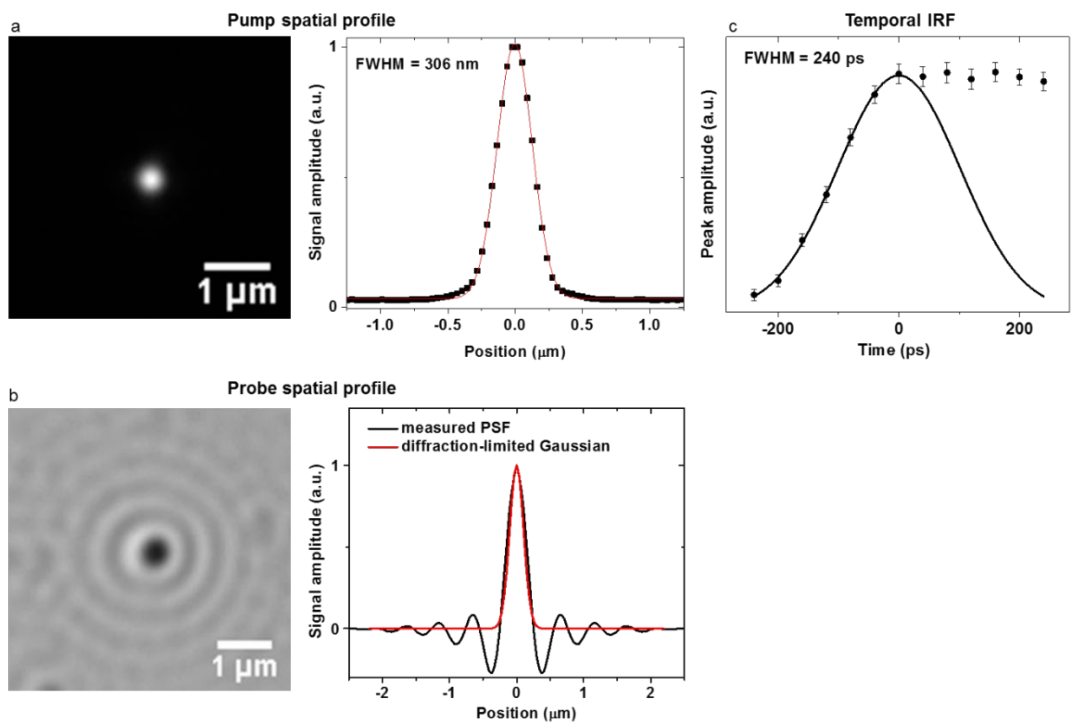

Figure S2. (a) pump reflection from a glass substrate imaged on the CMOS camera, exhibiting a Gaussian profile with a FWHM of 306 nm. (b) widefield probe PSF imaged using a 40 nm gold nanoparticle. The central part of the PSF corresponds closely to the expected diffraction-limited Gaussian. (c) Gaussian peak amplitude as a function of pump-probe time delay at early times for a stroboSCAT experiment on TIPS-Pentacene. The instrument response function of the system is estimated to be ~240 ps.

As shown in Figure S2, the 440 nm pump beam profile has a FWHM of 306 nm, which is ~ twice the diameter of a diffraction-limited spot. The total system spatial resolution is ~ $\sqrt{306^2 + (635/2.8)^2} = 381$ nm. Using a diffraction-limited pump could improve the resolution to a best-case scenario (with these wavelengths) of ~276 nm. We opted for a non-diffraction-limited beam by underfilling the objective in order to avoid polarization scrambling in the focal plane. The measured probe PSF (Figure S2b) is used in the simulations in Section 8.

The system temporal instrument response function (IRF) is determined to be ~240 ps using a half-Gaussian to fit the signal rise-time in a stroboSCAT experiment on TIPS-pentacene (Figure S2c).

## 6. Distinguishing scattering from normal reflection

It is possible to separate scattered and illumination light near the back focal plane of the objective by making use of the fact that small scatterers (e.g. individual energy carriers, quantum dots, nanoparticles etc.) near a refractive index interface radiate the majority of photons at the critical angle determined by the interface[4,5,28]. This directionality applies not only to elastic scattering but also to other light-matter interactions such as absorption, photoluminescence and inelastic scattering from deeply subwavelength particles near the refractive index interface. Here, we assume that the light traveling at the critical angle is due to elastic scattering. This phenomenon results in the scattered field being distributed in directions primarily associated with a high numerical aperture of the objective[4]. Since widefield illumination only requires a very low numerical aperture, the spatial frequencies of the scattered light and illumination beam are well separated near the back aperture of the objective. Indeed, it has been shown in iSCAT experiments that the illumination beam can be attenuated with an appropriate partial reflector while transmitting the vast majority of the scattered field in order to increase the iSCAT contrast significantly[4,5]. Similarly, the normally-reflected light can be entirely blocked, resulting in a sensitive darkfield backscattering microscope[2].

To verify that the same separation of spatial frequencies is present in stroboSCAT, and to confirm the signal source is scattering from a collection of small particles (and not just a change in the reflectivity $r$ of the interface), we introduced an aperture near the back aperture of the objective (Figure S1) to be able to interchange between stroboSCAT and normal reflectivity modes of imaging.

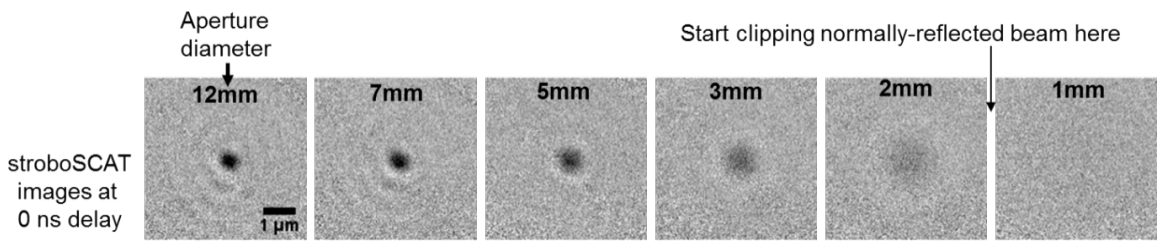

Figure S3. stroboSCAT contrast for TIPS-pentacene at 0 ns delay. The images from left to right show how the contrast changes as high spatial frequencies of the scattering field are progressively filtered out using an aperture near the back aperture of the objective.

Figure S3 shows the effect of spatially filtering the scattered field only while the normally-reflected illumination field is fully transmitted. The stroboSCAT signal magnitude reduces and the spatial extent of the signal on the detector plane increases as high spatial frequencies are filtered out. These experiments indicate that the scattered field emitted toward high-NA associated directions (from a collection of sub-wavelength-sized objects) is indeed responsible for the observed signal.

Furthermore, a 3 mm beam stop attached to a thin wire was inserted to block the normally-reflected beam while transmitting the scattered field only[2]. Using the same experimental conditions, we were unable to observe any signal above noise on the detector. In contrast, dust particles or other strongly-scattering objects switch from dark contrast on a bright background to bright contrast on dark background, as expected for darkfield microscopy.

The same control measurements were performed on all samples, giving identical results. Taken together, these observations indicate that the interferometric cross-term dominates the differential signal magnitude, i.e. $s^2 \ll rs\cos\varphi$, and the reflectivity $r$ does not change significantly between pumped and unpumped states. Thus, darkfield microscopy or spatially-filtered transient reflection are not able to achieve the same sensitivities as stroboSCAT.

## 7. Supporting experimental data

All reported injected energy carrier densities, $n_0$, are calculated as $n_0 = j\alpha$, where $j$ is the peak pump fluence in photons/cm$^2$, and $\alpha$ is the absorption coefficient reported in the literature. Peak photon fluence is calculated from peak energy fluence, which is defined here as $2E/\pi r^2$, with $E$ being the pulse energy and $r$ being the beam radius at $1/e^2$. For each sample, a pump-power dependence over several time delays is performed in order to ensure that the rate of decay of the stroboSCAT signal peak amplitude is power-independent over the range of powers used. In this way we ensure that many-body effects such as Auger recombination do not contribute significantly to the determined diffusivities. Using higher powers may lead to incorrect estimations of the diffusion coefficient as the population distributions approach flat-top profiles rather than Gaussian profiles, which, if fit with a Gaussian function, will appear as erroneously large distribution widths[26]. For TIPS-Pentacene, power-dependent behavior is non-trivial and discussed in more detail in Figure S11.

### 7.1. Silicon

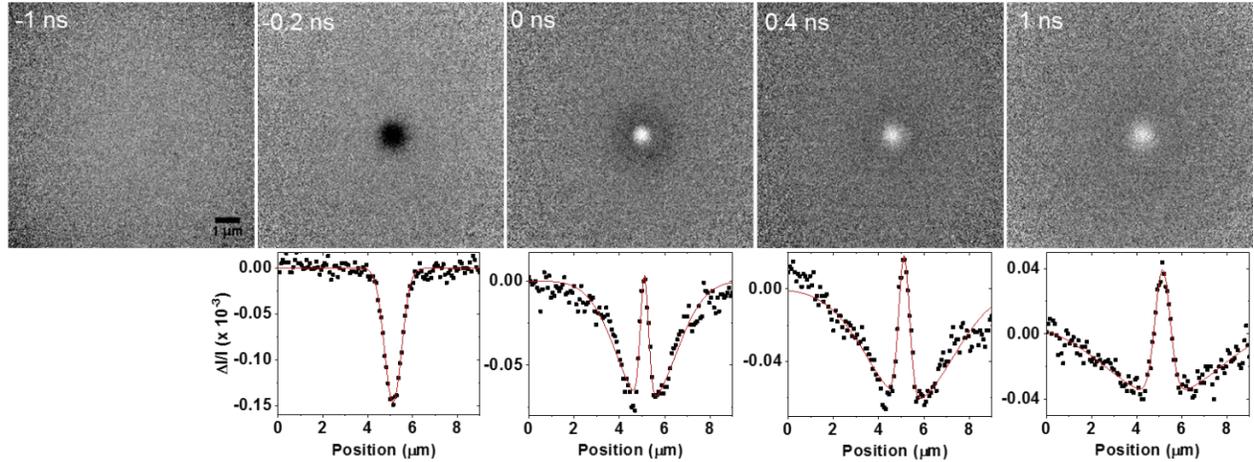

Figure S4. stroboSCAT time series on a p-doped Si wafer. Just before 0 ns pump-probe delay, the signal exhibits negative (dark) contrast, which rapidly expands with time. At 0 ns pump-probe delay, a positive-going contrast appears in the center of the distribution, indicating a different species is formed rapidly. Double-Gaussian fitting of the signal allows extracting the diffusivities for each species. The negative Gaussian expands at a rate of 35 ± 8 cm$^2$/s and the positive Gaussian expands at a rate of 0.6 ± 0.2 cm$^2$/s (Figure 1). These are close to reported values for electron[29] and thermal[30] diffusivities in Si. In addition, heat and free electrons are indeed expected to produce opposite changes to the refractive index of silicon at optical wavelengths[31,32]. We therefore attribute the negative (dark) contrast to photogenerated electrons, and the positive (bright) contrast to heat deposited in the lattice through relaxation of excess energy from above-bandgap excitation as well as electron-electron, electron-hole and electron-phonon scattering that happen primarily in the center of the carrier distribution, where the carrier density is largest. Note that this is the only sample where fitting errors are larger than the variation from dataset to dataset, primarily due to fast-diffusing carriers, low contrast and the need for double-Gaussian fitting. Fluence is 0.5 mJ/cm$^2$ ($n_0 \approx$ 2 x 10$^{19}$ cm$^{-3}$)[33]. 10-scan averaging and 8-pixel wide integration are used for the plots shown. Both pump (440 nm) and probe (640 nm) are far above the silicon bandgap (~1100 nm) and distinct from pump-induced absorption changes.

## 7.2. CsPbBr$_3$

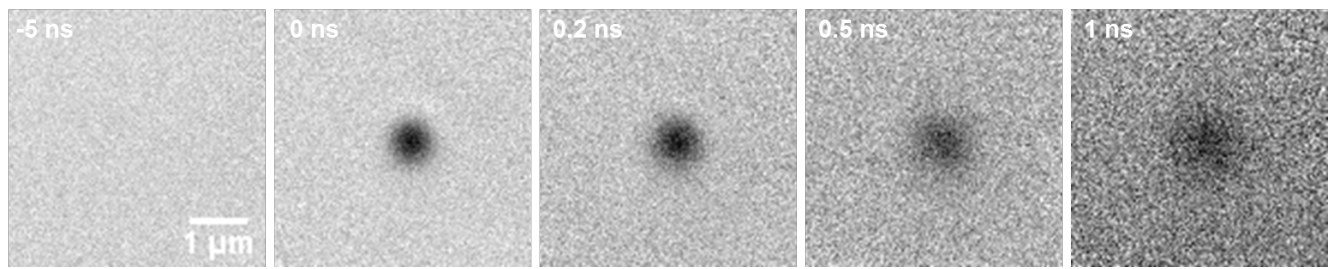

Figure S5. stroboSCAT time series on a CsPbBr$_3$ single crystal. The extracted diffusivity from this dataset is $1.00 \pm 0.08$ cm$^2$/s (Figure 1). The average and standard deviation across 6 datasets and 2 crystals are $1.0 \pm 0.2$ cm$^2$/s. Fluence is 21 µJ/cm$^2$ ($n_0 \approx 4 \times 10^{18}$ cm$^{-3}$)[34]. The pump at 440 nm is above bandgap, while the probe at 640 nm is non-resonant with both ground and excited state absorption.

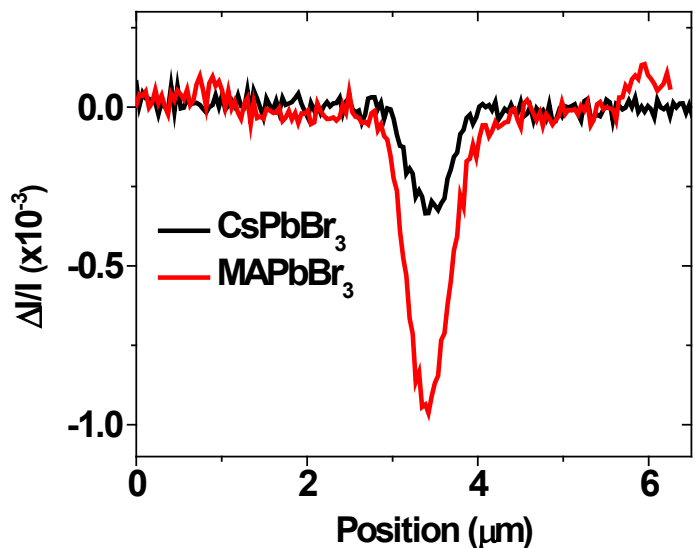

Figure S6. stroboSCAT contrast difference at 0 ns pump-probe delay at a fluence of 10 µJ/cm$^2$ for MAPbBr$_3$ vs. CsPbBr$_3$ single crystals. The contrast magnitude is lower by a factor of 2.7 in CsPbBr$_3$ despite both systems having similar absorption coefficients and both being probed with below-bandgap, off-resonant light. The extent to which contrast magnitude can be used to gain information on electron-phonon coupling in these different systems will be the subject of future investigations.

## 7.3. MAPbBr$_3$ polycrystalline films

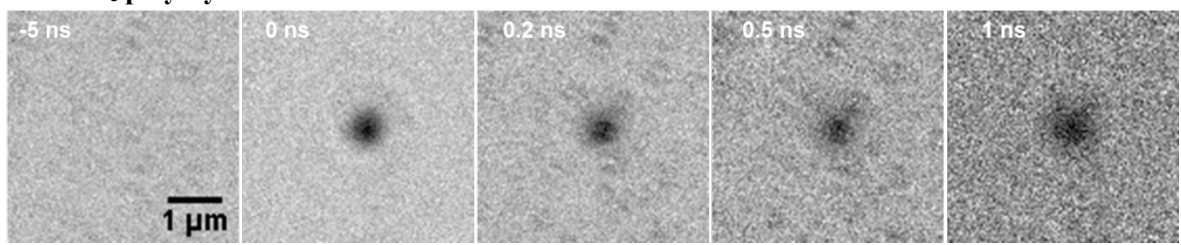

Figure S7. stroboSCAT time series on a MAPbBr$_3$ polycrystalline film. Note that background contributions due to sample vibrations/drift during acquisition are larger for heterogeneous samples than for samples that scatter homogeneously. The extracted diffusivity from this dataset is 0.15 ± 0.02 cm$^2$/s. The average and standard deviation across 5 datasets are 0.16 ± 0.05 cm$^2$/s. Fluence is 10 µJ/cm$^2$ ($n_0 \approx$ 2 x 10$^{18}$ cm$^{-3}$)[35]. The pump at 440 nm is above bandgap, while the probe at 640 nm is non-resonant with both ground and excited state absorption.

## 7.4. TIPS-Pentacene

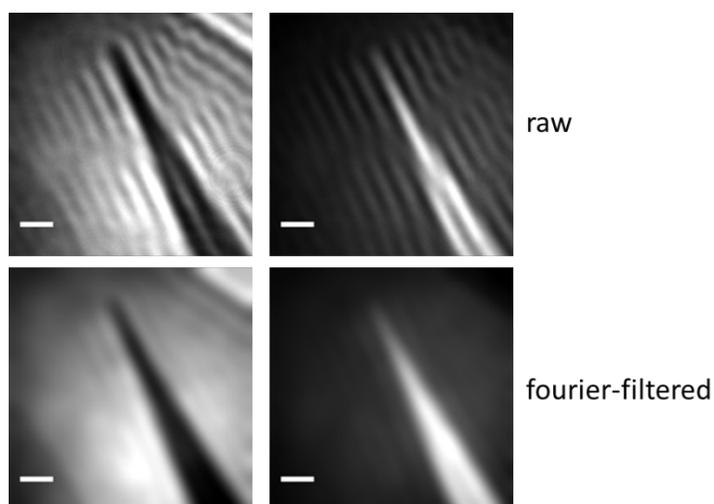

Figure S8. The iSCAT images of TIPS-pentacene crystalline domains shown in Figures 2b of the main text are filtered using a Fourier bandpass, allowing to remove stripes from probe light diffraction off of crystalline interfaces in the sample. The diffraction stripes are a consequence of using widefield illumination. The raw images are shown here at the top. Scale bars are 1 µm.

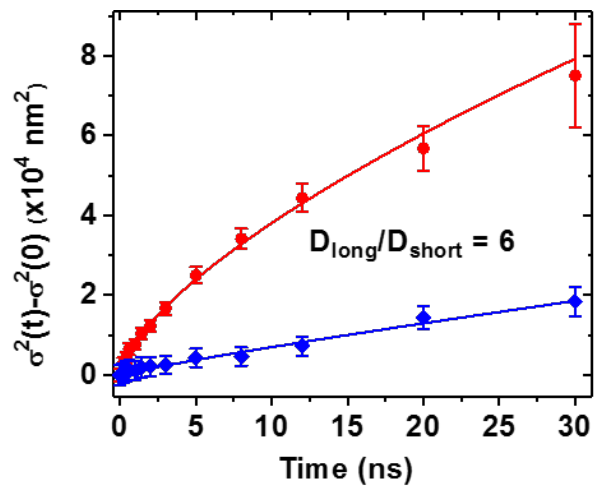

Figure S9. Anisotropic intra-domain diffusivity in TIPS-Pentacene. Diffusivity along the long axis (red) is approximately 6 times larger than along the short axis (blue), as expected for anisotropic π-stacked systems[22,36].

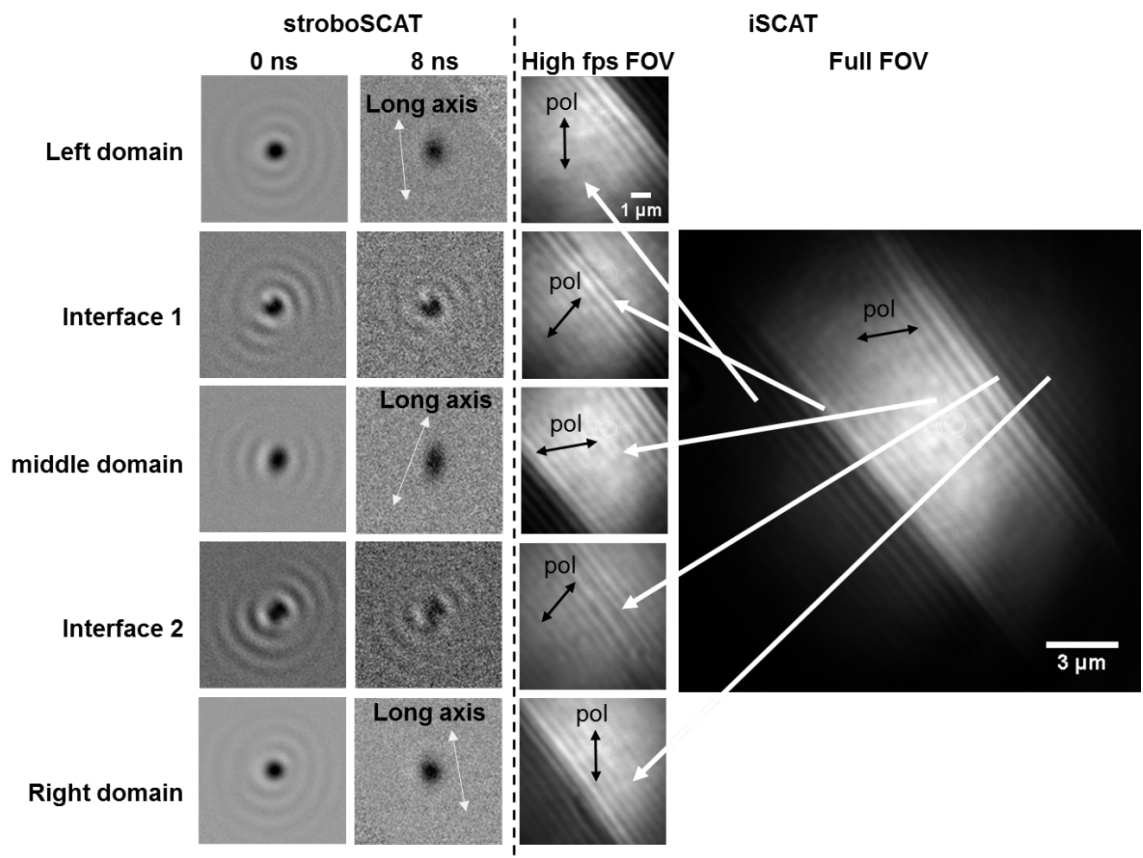

Figure S10. Additional datasets on TIPS-Pentacene in three crystalline domains and at two GBs. stroboSCAT images are shown on the left at 0 and 8 ns time delays. In the middle column, corresponding iSCAT (pump OFF) images at the probe polarizations used for each dataset are shown. On the right, a larger field-of-view image is used to capture the three domains in a single image at a given polarization – note that the contrast for each crystalline domain in iSCAT (pump OFF) images switches between bright and dark depending on the probe polarization used. The middle domain appears somewhat out-of-plane (i.e. the *c*-axis is not perpendicular to the substrate plane), as evidenced by a non-circular profile of the excitation spot at 0 ns in the stroboSCAT image, which we attribute to projection distortion. This non-planar orientation of the crystalline domain was observed in around 20% of the domains investigated in these films. For these datasets and those shown in the text, the pump was circularly polarized. When investigating GBs, the probe polarization is set to equalize contrast across domains, as shown explicitly in the iSCAT images above. We tested the validity of this approach by verifying that similar results were obtained with a circularly-polarized probe both in domains and at GBs, although the signal-to-noise ratio was poorer when using a circularly polarized instead of linearly polarized probe beam.

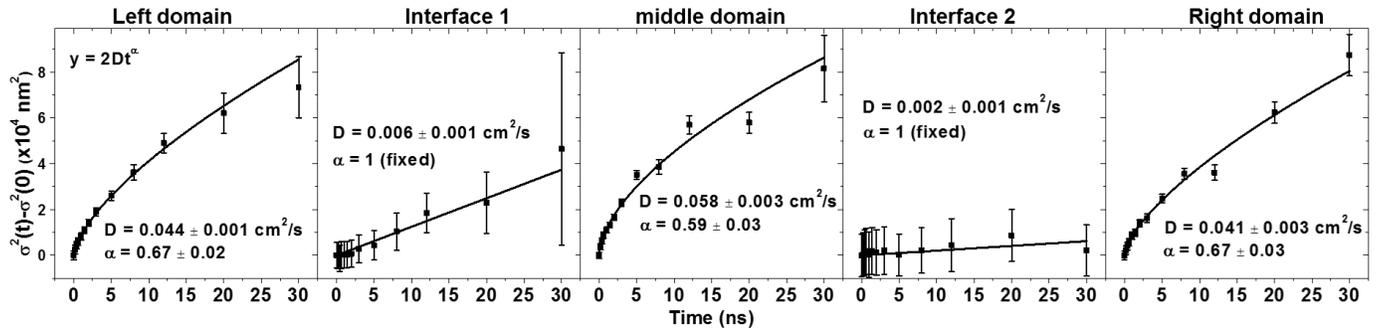

Figure S11. Population expansion plots for the data in Figure S10. A subdiffusive model is used to fit the data and extract an effective $D_0$ and to compare among domains and GBs. Due to the larger noise at GBs and apparently linear diffusive behavior, we used a linear model there instead to avoid over-fitting.

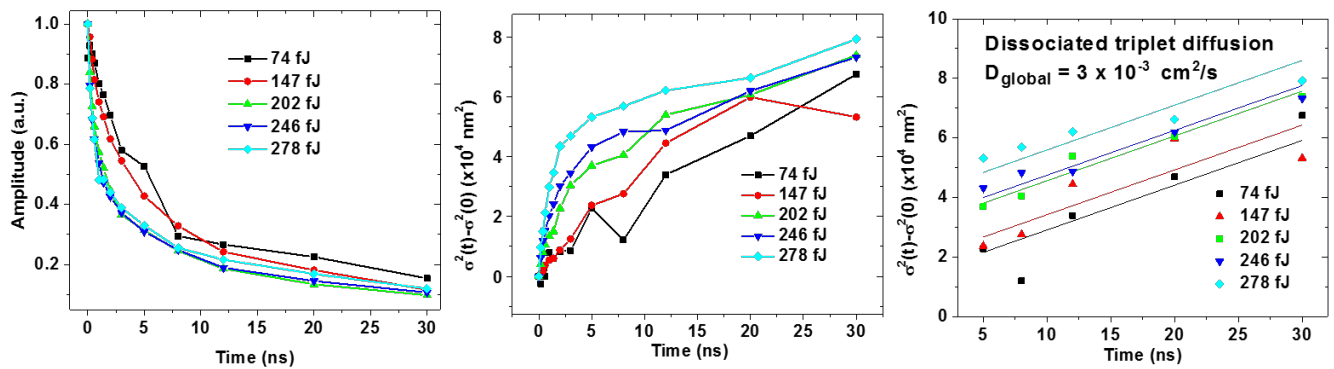

Figure S12. In TIPS-Pentacene, early-time dynamics are dictated by singlet-fission into two triplets and the reverse process of triplet fusion into a singlet, which affects the rate of exciton migration since singlets diffuse faster than triplet excitons[37,38]. The combination of this interplay with singlet-singlet annihilation gives rise to nonlinear diffusive dynamics for as long as triplets are strongly bound – up to nanoseconds[39], as seen in the middle panel. To eliminate singlet-singlet and free triplet-triplet annihilation and concentrate on intrinsic singlet-triplet interchange, our measurements are taken at 147 fJ (140 uJ/cm$^2$ peak fluence), as we detect no significant changes in the peak amplitude decay between 74 and 147 fJ (the left panel plots the normalized peak Gaussian amplitudes of the stroboSCAT signal). At higher fluences, faster decay of the peak amplitudes at early times is observed, resulting in faster apparent diffusion. At long times (>5 ns), once all triplets are fully separated, linear diffusivities of 3 x 10$^{-3}$ cm$^2$/s are observed for all fluences used (right panel), which we assign to free triplet diffusion in TIPS-pentacene.

## 7.5. MAPbI$_3$

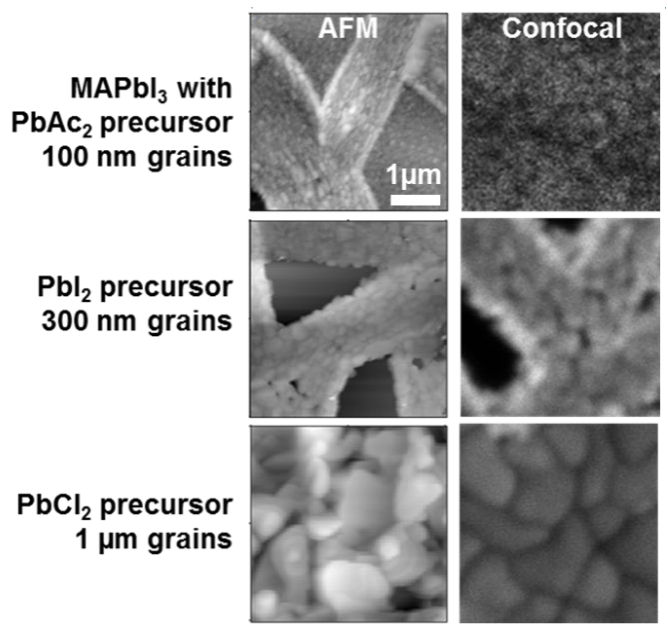

Figure S13. Example AFM and confocal images for the three MAPbI$_3$ films studied, allowing to extract approximate domain sizes for each sample. AFM and confocal images are not from the same sample areas.

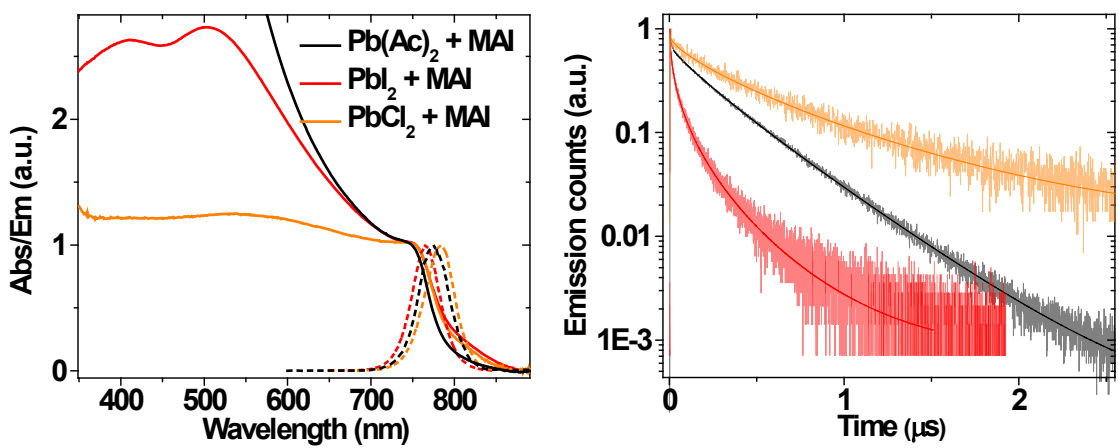

Figure S14. Absorption spectra, emission spectra, and emission lifetime for MAPbI$_3$ films prepared from Pb(Ac)$_2$ (black), PbI$_2$ (red), and PbCl$_2$ (orange) precursors. The absorption spectra are baselined by subtracting the transmission changes due to scattering and reflectance to facilitate comparison between films. Emission spectra and lifetimes are measured using 470 nm excitation at ~0.1 µJ/cm$^2$ on a FluoTime 300 instrument (PicoQuant). The curves overlaid on emission decay traces correspond to stretched exponential tail fits $A \times \exp(-t/\tau)^\beta$ with parameters $\tau$ = 276, 24, 379 ns and $\beta$ = 0.88, 0.48, 0.74 for films made from Pb(Ac)$_2$, PbI$_2$, and PbCl$_2$ precursors, respectively. The decay lifetimes correspond closely to those reported for related preparation protocols[27,40,41].

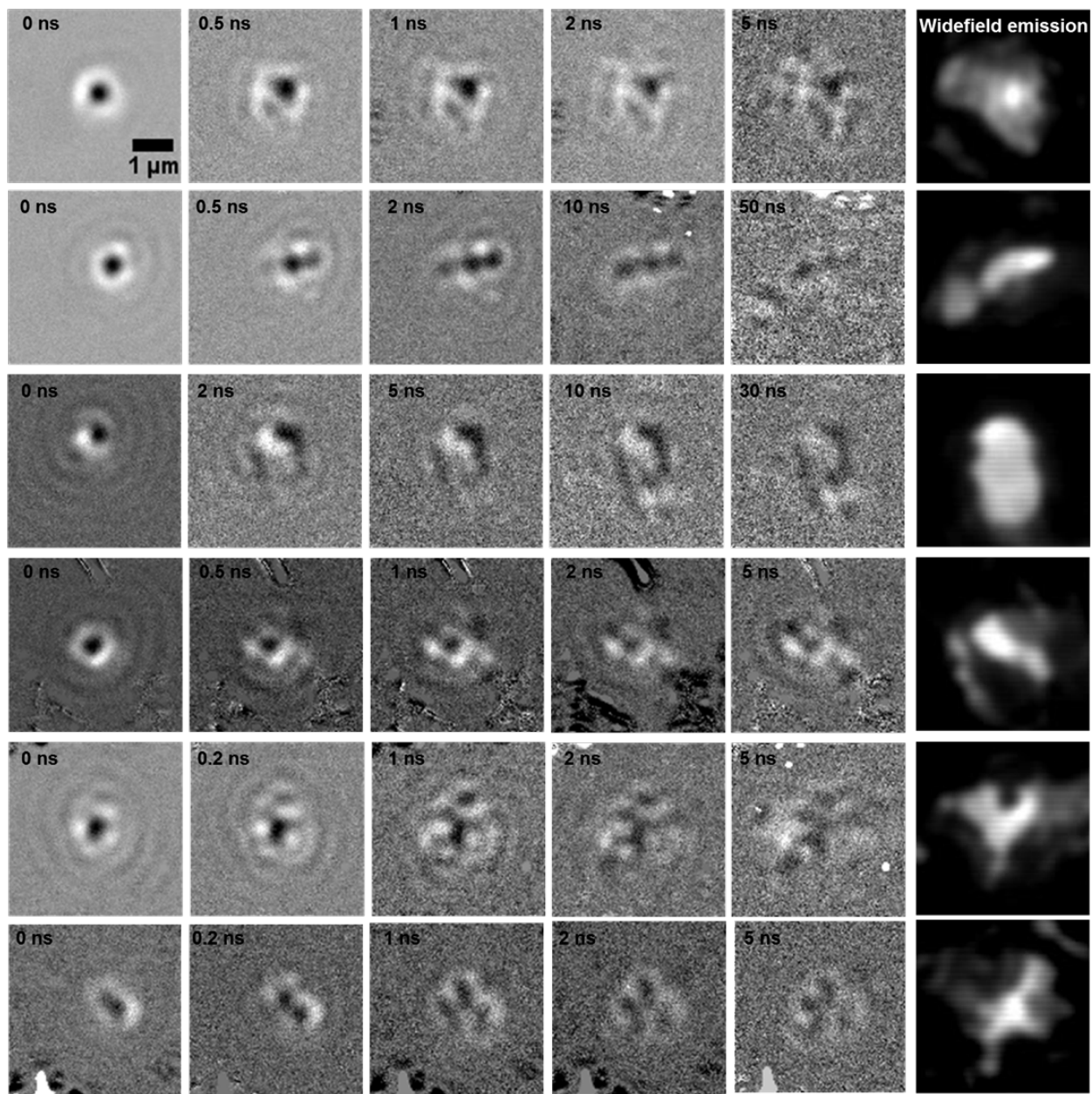

Figure S15. Illustration of spot-to-spot heterogeneity within a single film of MAPbI$_3$ made with PbCl$_2$ precursor. Both the stroboSCAT time series and widefield emission show vastly different inter-domain connectivity patterns at each spot. Trapping times differ too, with some spots exhibiting population expansion up until 5 ns while others display expansion up until 50 ns. Good correspondence between widefield images and stroboSCAT images at late times confirms that stroboSCAT captures the full extent of carrier diffusion; however, some differences between widefield and stroboSCAT images at late times can arise if electron and hole transport are substantially different, for example due to preferential trapping of one species at MBs, since excess free carriers of one species will not give rise to photoluminescence, but will still change the polarizability of the medium. All excitation fluences are 10 µJ/cm$^2$ ($n_0 \approx 1.5 \times 10^{18}$ cm$^{-3}$)[42].

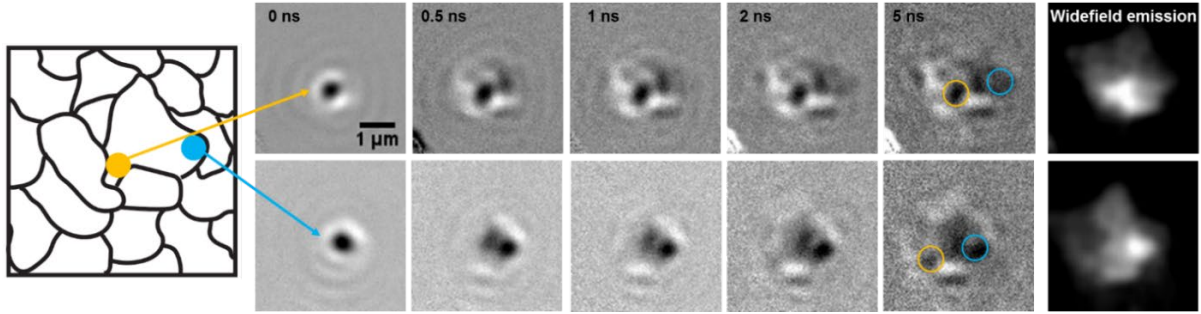

Figure S16. Specific intra- and inter-domain connectivity patterns determine carrier diffusion pathways in large-domain MAPbI$_3$ (PbCl$_2$) films. Here we show stroboSCAT time series and widefield emission images when pumping at two slightly offset spots in the same region of the film. The gold and blue circles show where the initial distribution of free carriers are injected. The same circles are overlaid on the 5 ns images to allow spatially correlating the two images. The final carrier population distribution and contrast flips correspond almost perfectly, confirming that fast intra-domain transport and preferential inter-domain connectivities across depth- and lateral-dependent paths of least resistance through MBs define overall carrier transport pathways. Note that even though carriers can cross MBs more readily below the film surface, MBs still act as bottlenecks to carrier transport due to their average resistivities being higher than intradomain resistivities. Note also that some MBs clearly possess larger average resistivities than others, acting as more effective bottlenecks to carrier transport.

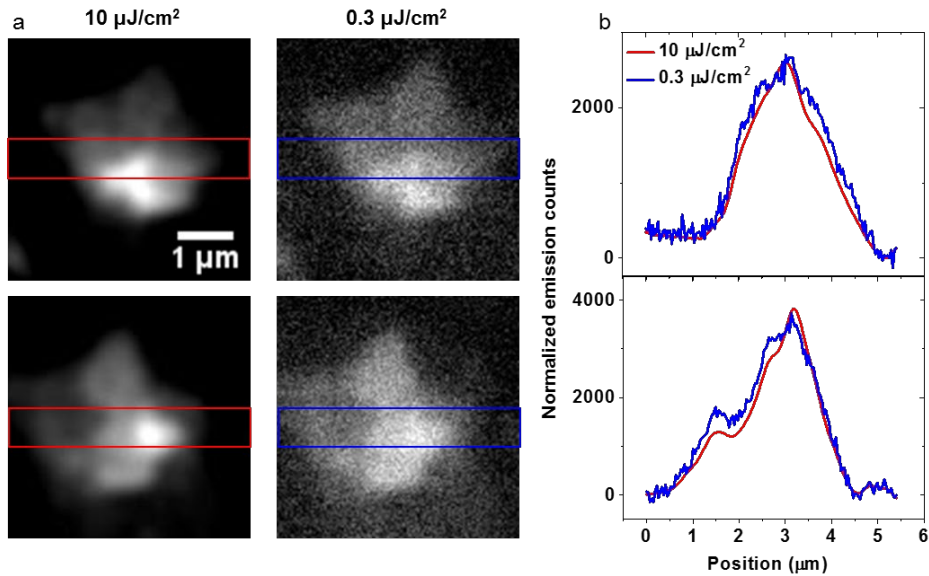

Figure S17. Pump-power-dependent widefield photoluminescence profiles in MAPbI$_3$(Cl) films, excited at the two different spots from Figure S16. Red and blue curves in (b) correspond to fluence-normalized, integrated photoluminescence intensity profiles delimited by the rectangles in (a). Results show a small, ~10% variation of carrier population distribution on average despite a 30-fold reduction in peak fluence. Lower excitation powers display slightly broader distribution profiles, likely an indication that slightly fewer carrier-carrier scattering events impede diffusion at early times in the low-fluence measurements. The small deviation confirms that high-order recombination terms do not contribute substantially to the stroboSCAT measurements at the fluences used, and thus that the measurements are performed within a linear excitation regime.

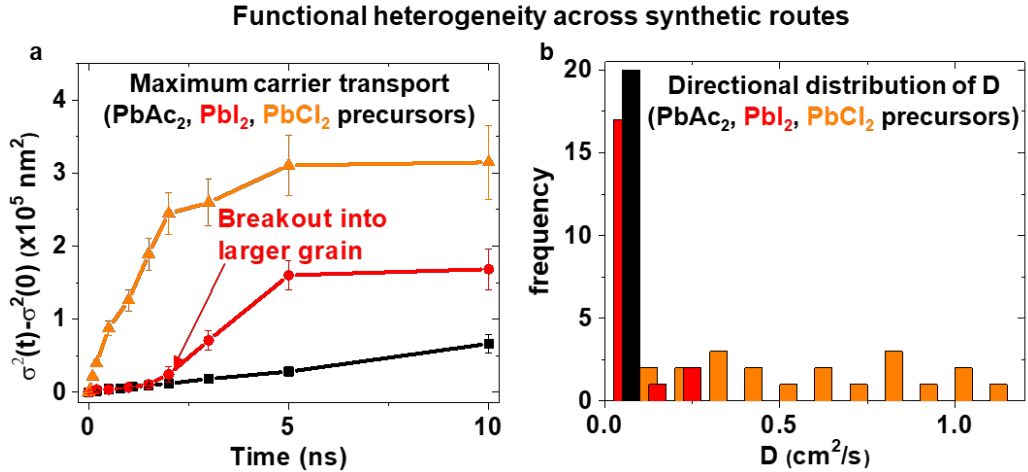

Figure S18. (a) Population expansion for the fastest azimuthal trajectory for each dataset from **Figure 3a**. Error bars represent the standard deviation of the fits to several experimental datasets taken at the same location in the films. (b) Distribution of diffusivities over 20 radial trajectories for perovskite samples prepared from three different precursor solutions.

## 8. *In situ* spectral interferometry on MAPbI$_3$(Cl) films

Several research groups have shown that spectral interferometry near an absorption resonance can be used to detect both the phase and amplitude response of deeply subwavelength nano-objects (e.g. single molecules and quantum dots)[43–46]. Indeed, with spectral interferometry, small phase shifts between the signal and reference fields result in dramatic changes to the spectral profile of the object under study near its absorption resonance. As such, using the available phase information, several studies reproduced with high certainty and accuracy the known depth at which these single nano-objects were located in the sample plane[43,44,46]. Building on these studies, we incorporated the ability to obtain spectral interferometry measurements in parallel to stroboSCAT. Whereas stroboSCAT provides the ability to track carrier motion in 3D, spectral interferometry provides complementary phase-sensitive spectral information with diffraction-limited resolution in the same field of view. The latter allows benchmarking stroboSCAT's phase sensitivity and helps reinforce our interpretation of depth-dependent transport across MBs.

We spectrally resolve the photoinduced signal in MAPbI$_3$(Cl) films close to the band edge in diffraction-limited volumes that allow distinguishing intra-domain vs MB signals. Previous spectral interferometry studies on single quantum dots or molecules showed that, depending on the depth of the object with respect to the reflective interface, the band edge spectral profile shows distinctive features with positive, negative or dispersive line shapes[43,44,47]. Similarly, the transient reflectance spectral response at the band edge of MAPbI$_3$ should show such distinct features if the photoinduced changes emanate from surface vs. subsurface free carriers.

Figure S19 shows representative spectral interferometry data from within MAPbI$_3$ domains vs. at MBs. Within domains, the observed dispersive line shape corresponds closely to that reported in the literature from bulk (non-spatially-resolved) transient reflectance spectroscopy of MAPbI$_3$ single crystals.[15] The spectral response at MBs (red traces) occur at the same spectral positions but are inverted. The exact spectral profile at the MB can vary somewhat from MB to MB, though it always displays a phase shift with respect to the intra-domain signal. Out of ~50 MBs we measured in two different films, ~40 of the spectra at MBs possess an inverted dispersive lineshape like the red trace shown in the left panel of Figure S19; others exhibit lineshapes that are halfway between negative and dispersive (right panel, red trace), indicating phase shifts between π/2 and π. These different MB profiles suggest that each MB has slightly different depth-dependent resistive properties, but that all MBs appear to be highly resistive at the film surface, one of the key scientific findings of our work. We also note that the small probe volumes employed here allow us to confirm that our interpretation of stroboSCAT results (which employ a widefield probe) is not compromised by interference effects arising from a more complex point spread function (i.e. from scattering sources at different locations).

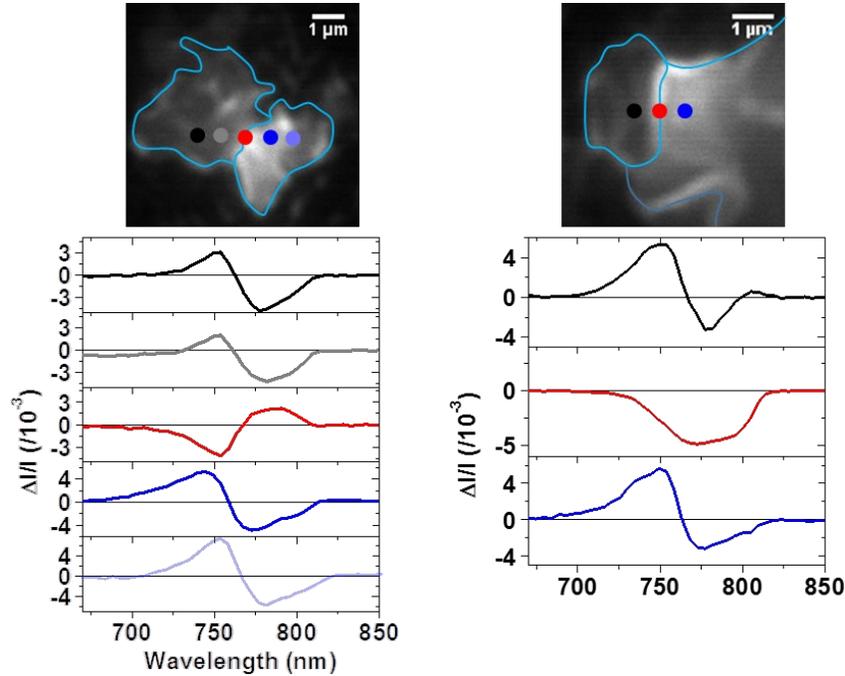

Figure S19. Spectral interferometry of intra-domain vs. MB photoinduced signals at a pump-probe time delay of ~ 0 ns. The images at the top are widefield emission images indicating the domains and boundaries being probed. The colored dots represent the overlapped pump and probe positions, with corresponding spectral profiles displayed below for each of these positions. Approximately 80% of measured MBs show dispersive lineshapes like that found in the left panel, while the rest display lineshapes between dispersive and negative, like that shown in the right panel.

The pump and probe spots can be displaced with respect to one another to follow the migration of carriers from one domain to another through a MB. Figure S20 displays the spectral and temporal dependence of the signal for three probe positions: one spatially overlapped with the pump (position represented by the blue dot) within a domain (black), one away from the pump in an adjacent domain (gold), and one at the domain boundary separating the two domains (red). The spectral phase within the two domains is identical, but the signature spectral phase flip is again observed at the MB, confirming that the transport pathway between the two domains is below the MB surface. The temporal dependence of the signal follows the expected kinetic behavior: the arrival time of the signal (i.e. the carriers) is delayed at the MB and even further delayed in the adjacent domain, confirming that a significant population of carriers from the pumped domain does cross through the MB into the adjacent domain, but is somewhat slowed down by the MB. The recovery of the original intra-domain phase once carriers cross into the adjacent domain indicates that carriers rapidly diffuse in 3D once away from the MB, as we had originally inferred from stroboSCAT. Overall, by relating these phase-sensitive spectro-temporal data at key specific locations to what we have observed in spatio-temporal stroboSCAT, we not only confirm our original interpretation that transport through MBs occurs through sub-surface pathways, but also rule out potential artefacts in both spectral interferometry and stroboSCAT that could arise due to strong pump scattering off of the MB (avoided here because the pump is spatially separated from the MB), probe interference effects from back-surface reflections (avoided here because our probe is tightly focused on the front surface of the film), or spectral changes as carriers transport through the material that could give rise to fluke signals at our original probe wavelength of 640 nm.

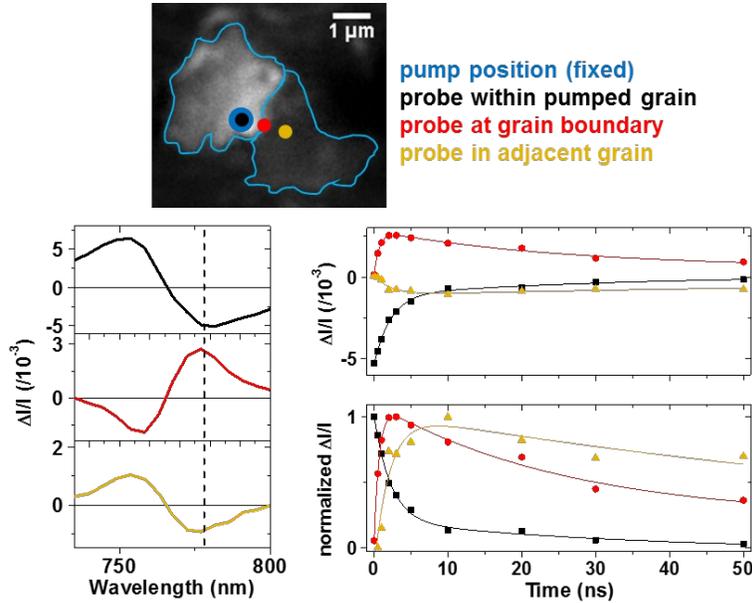

Figure S20. Spectral and temporal dependence of the transient scattering signal for three probe positions: one overlapped with the pump within a domain (black), one away from the pump in an adjacent domain (gold), and one at the domain boundary separating the two domains (red). The kinetic traces shown are taken at the spectral position indicated by the dashed line in the bottom left panel.

9. **Simulations of depth-dependent carrier diffusion in polycrystalline films using the finite element method**

We simulate depth-dependent charge diffusion in $MAPbI_3(Cl)$ films using finite element analysis implemented in the MATLAB PDE toolbox. The parabolic diffusion equation is solved in a heterogeneous environment consisting of domains separated by abrupt domain boundaries. The code is implemented in a two-dimensional *x-z* (lateral-axial) plane. In brief, the simulations quantitatively reproduce the observed diffusion behavior and associated stroboSCAT contrast. The key aspect of the simulations is the inclusion of a depth-dependent diffusion coefficient at MBs. The model depends most sensitively on the depth at which the diffusion coefficient increases from ~0 at the surface to its intra-domain value. All simulation parameters are detailed in Section 8.1, and the results are further supported by additional datasets and simulations in Section 8.2. This last Section includes structural information correlated to stroboSCAT measurements in the same field of view to further constrain the model and confirm our interpretation: contrast flips in the stroboSCAT data on $MAPbI_3(Cl)$ films occur specifically at MBs and indicate that when carriers encounter MBs, they cannot cross into neighboring domains at the film surface, instead moving deeper into the film, where MBs are not as resistive.

## 9.1 Simulation parameters and assumptions

The MAPbI$_3$(Cl) film thickness is 300 nm as measured by AFM. MBs are assumed to lie approximately perpendicular to the substrate plane. The films are assumed to be 1 domain thick (i.e. no MBs parallel to the substrate plane), as illustrated in Figure 3d of the main text. These assumptions are reasonable based on cross-sectional SEM measurements taken on MAPbI$_3$(Cl) films prepared using the same protocol[48]. The MBs are simulated as 200 nm thick. We choose 200 nm based on confocal fluorescence microscopy measurements that show occasionally completely dark MBs even though the image is convolved with a ~ 200-300 nm point spread function, implying that the effect of MBs is felt over a >150 nm region despite the fact that they may be far thinner. For simulations with correlated structural measurements (Section 8.2), slight adjustments to the MB thickness are made in the simulation if they are determined to be larger than the diffraction limit in the confocal measurement. Since simulations are performed in the *x-z* plane, MBs that are not perfectly in the *y-z* plane are also simulated with larger thicknesses. Von Neumann boundary conditions are assumed at all film edges, though the proportion of carriers reaching the lateral edges are negligible over the simulation time. Interparticle interactions are ignored, supported by the fluence-dependent measurements shown in Figure S17. A first-order rate constant for recombination is included (see below). The starting condition is an injected carrier density profile from the pump with a FWHM of 306 nm and exponentially decaying as a function of depth with a *1/e* penetration depth of 60 nm.

The key simulation parameters are shown in Figure S21. First, the MBs are assumed to have depth-dependent resistive properties. This assumption is based on conductive-probe AFM measurements that show that the surface of GBs in MAPbI$_3$ films are infinitely resistive, but that carriers do cross domain boundaries, leading the authors to postulate that the resistivity of GBs decreases as a function of depth[49]. While the depth-dependent resistive profiles can adopt many forms, and may be GB-dependent and film-dependent, we adopt the simple form shown in Figure S21a: an infinitely-resistive surface as previously measured using AFM, with a rapid drop-off to a value similar to the bulk resistivity. Note that the value that is varied in the simulations is the diffusion coefficient, related to resistivity $\rho$ through the Einstein equation, $D=kT/(q^2 N\rho)$, for charge $q$ and charge number density $N$. The simulations qualitatively reproduce the data over a wide range of drop-off rate and final value for the diffusivity, but the onset of the drop-off is well-constrained: carriers need to pass through the MB in the region of 40-80 nm below the surface to reproduce the signals we observe. Importantly, including depth-dependent resistive profiles at MBs is crucial to reproducing stroboSCAT data for every simulated dataset – the phase flips cannot be reproduced with any other parameters described herein, and we have never observed phase flips due to changes in the carrier density in these materials over 4 orders of magnitude in pump fluence.

Second, the recombination rate at the top and bottom surface of the films are assumed to be 10 times larger than in the bulk, based on multiple studies showing that surface recombination dictates carrier lifetimes in MAPbI$_3$ perovskites[15,50]. The recombination rate profile shown in Figure S21b is determined based on the knowledge that surface contributions dominate photoluminescence decay with 400 nm excitation (~40-60 nm penetration depth)[50]. The simulation reproduces the experimental results over a wide range of parameters for the recombination rate profile vs depth.

Finally, after simulation carrier diffusion, the stroboSCAT contrast is reproduced by applying a contrast scaling term based on the phase sensitivity of iSCAT (see Section S2). A $\pi$ phase shift, which occurs every $\lambda/4n \approx 60$ nm, corresponds to a contrast flip from negative to positive. To account for the finite probe penetration depth of $d = 67$ nm, and the fact that probe photons that reach the detector upon scattering at a sample depth $z$ pass through $2z$ of the material, we attenuate the simulated incident probe light using a damped cosine function $-\cos(\Delta\phi) * e^{-2z/d}$, where $\Delta\phi$ is the phase shift $\Delta\phi = 4\pi n z/\lambda$. The negative sign accounts for the fact that the phase at the surface is $-\pi$ due to the Gouy phase (Section 2). Finally, once this scaling is applied, the contrast is integrated across the depth of the film to simulate the stroboSCAT signal, which is then convolved with the probe PSF (Figure S2b).

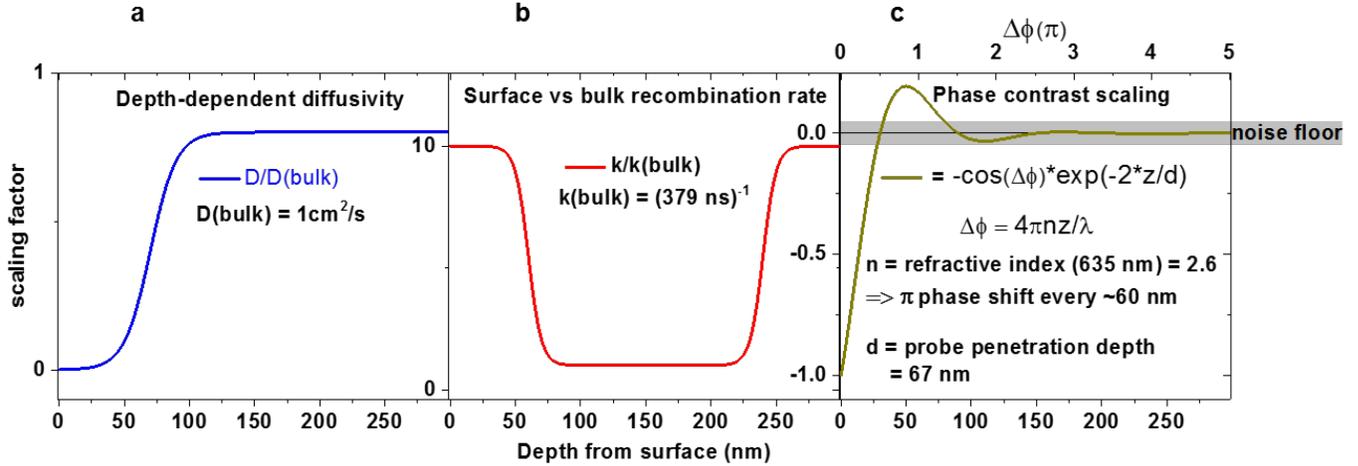

Figure S21. Depth-dependent simulation parameters: diffusivity (a), recombination rate (b), and contrast scaling for a probe wavelength of 635 nm in MAPbI$_3$ due to interferometric phase sensitivity (c). The noise floor for the modeled experiments is indicated by the grey box in (c). See text in Section 8.1 for details.

The calculated carrier distributions prior to convolution with the imaging system's PSF and contrast scaling for the data in Figure 3a of the main text are shown in Figure S22.

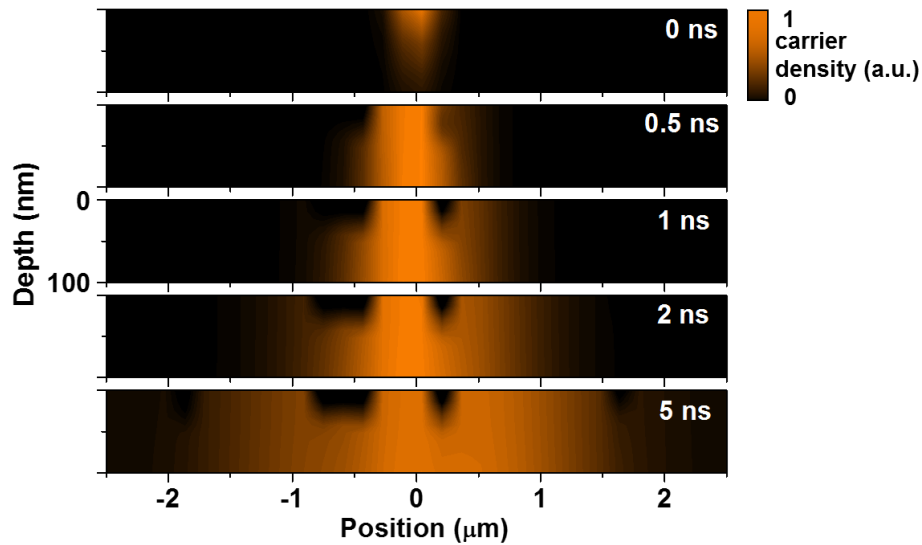

Figure S22. Simulated carrier distributions in the *x-z* plane for the PbCl$_2$ precursor film for the data shown in Figure 3a. The simulations use a 300 nm thick film, as measured by AFM; only the top 100 nm is displayed here, which is more representative of the actual probing depth in our experiments.

### 9.2 Simulation results with structural correlation: confirming stroboSCAT contrast flips occur at MBs.

To confirm that tracing carriers in 3D using stroboSCAT provides structural information, we undertook measurements on a well-defined region in a MAPbI$_3$(Cl) film with domains that are clearly visible and separated in confocal emission microscopy. The use of an optical method (rather than AFM or SEM) for structural correlation is necessary as the films need to be imaged on the same side as the probe is incident, i.e. through the sample substrate, and in oxygen- and moisture-free conditions. Figure S23 shows the region used for correlated measurements, with a well-defined central domain. A watershed algorithm in imageJ[7,8] is used to delimit the location of MBs. Although most of the literature refers to these MBs as grain boundaries, confocal emission microscopy or AFM in principle do not provide direct proof that the observed features are indeed GBs, which would require a microstructural crystalline-phase sensitive technique like electron backscattering diffraction (EBSD).

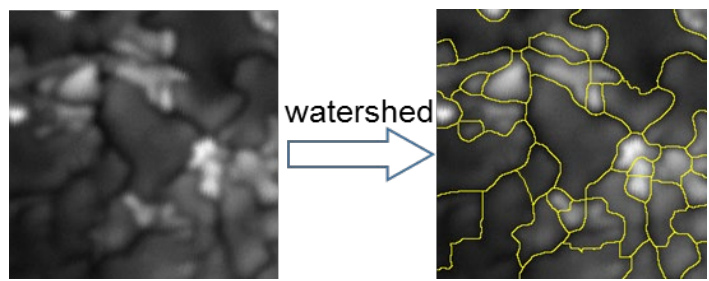

Figure S23. Region in a MAPbI$_3$(Cl) film used for structurally-correlated stroboSCAT measurements. A watershed algorithm is used to identify the position of domain boundaries. Domain boundaries in these films are dark in confocal emission microscopy, as has been previously observed[27]. Surface cracks would not give rise to large contrast changes in confocal emission unless they extend to depths comparable to or greater than the microscope's depth of focus, ~150-200 nm.

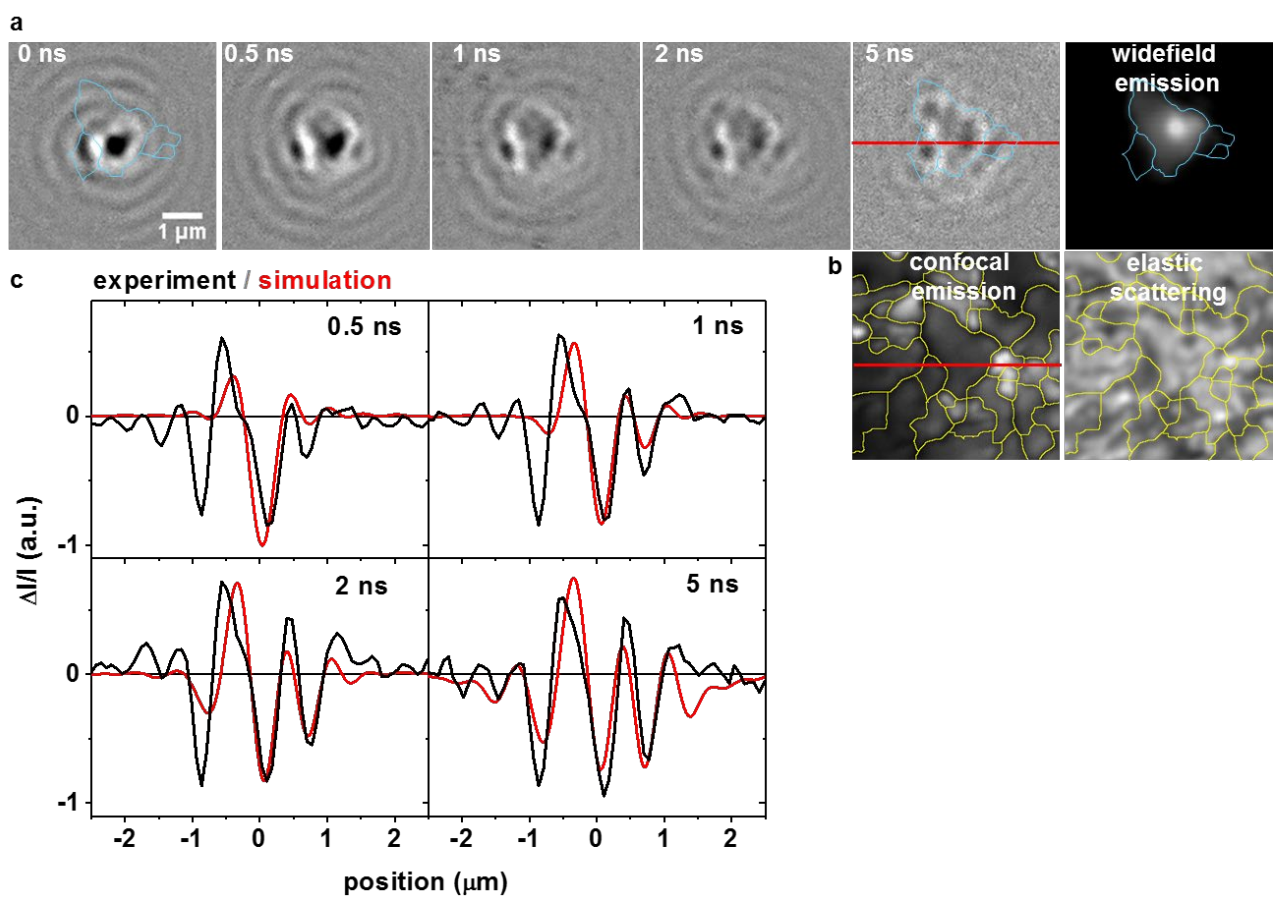

Figure S24. Structurally-correlated stroboSCAT-confocal fluorescence measurements. (a) stroboSCAT time series with correlated widefield emission from confocal excitation. The overlaid cyan curves show the 4 most relevant domains taken from Figure S23. (b) Correlated scanning-beam confocal fluorescence and elastic scattering at 635 nm. (c) Experimental and simulated stroboSCAT data for different time delays through the line cut indicated by the horizontal red line in (a) and (b). The simulations use the MB positions through the line determined from the confocal image in (b).

Figure S24a shows the stroboSCAT data when exciting the large central domain. By overlaying the MB positions determined from confocal microscopy on the 0 ns and 5 ns frames, it is clear that the positive (light) contrast in stroboSCAT arises at the MB locations. The same picture as that described in Figure 3 of the main text emerges: as carriers approach MBs, the stroboSCAT contrast flips from negative to positive, indicating that carriers pass MBs only at locations below the film surface. Once carriers have crossed MBs, they again freely move in 3 dimensions within domains, thus restoring the negative contrast. Since the stroboSCAT signal is strongest near the illuminated surface (Figure S21c), where the contrast is negative, stroboSCAT gives rise to negative contrast within domains and positive contrast at MBs, clearly delineating the structure of these polycrystalline films.

These deductions are confirmed with the finite element diffusion simulations shown in Figure S24c and Figure 3f, which reproduce the signal almost quantitatively despite using a relatively simple model. In the simulations shown in Figure S24c, the positions of the MBs are constrained to those determined in confocal microscopy.